\documentclass[%
 reprint,
 amsmath,amssymb,
 aps,
prb,
]{revtex4-2}

\usepackage{graphicx}
\usepackage{dcolumn}
\usepackage{bm}
\usepackage{slashed}
\usepackage{tikz}
\usepackage[compat=1.1.0]{tikz-feynman}
\usetikzlibrary{decorations.markings}
\usetikzlibrary{arrows.meta}
\usepackage{subcaption}
\usepackage{float}
\usepackage[markup=underlined]{changes}

\usepackage{ragged2e}
\DeclareCaptionJustification{justified}{\justifying}
\usepackage{braket}
\def\vk{\mathbf{k}}
\def\D{\mathcal{D}}
\def\e{\text{e}}
\def\d{\mathrm{d}}
\def\tr{\mathrm{Tr}}
\def\log{\mathrm{log}}
\def\i{\text{i}}
\def\S{\mathcal{S}}
\def\E{\text{E}}
\def\r{\text{r}}
\newcommand{\multiref}[2]{\ref{#1}--\ref{#2}}

\begin{document}

\preprint{APS/123-QED}

\title{Momentum space entanglement of four fermion field theory}

\author{Weijun Kong}
\affiliation{%
Department of Physics, Tsinghua University, Beijing 100084, P. R. China
}%


\author{Qing Wang}
\affiliation{
Department of Physics, Tsinghua University, Beijing 100084, P. R. China\\
Center for High Energy Physics, Tsinghua University, Beijing 100084, P. R. China
}%


\date{\today}

\begin{abstract}
Momentum space entanglement of four fermion field theory is calculated from the Wilsonian effective action pertubatively using replica trick, local terms in low energy effective action are proved to be non-relevant pertubatively and nonlocal terms are the only source of entanglement between different momentum modes. 
The final result can be represented by a set of basketball Feynman diagrams with new Feynman rules proposed to inteprete them.

\end{abstract}

\maketitle


\section{Introduction}
Entanglement entropy, a fundamental measure of quantum correlations, has emerged as a powerful conceptual tool across several frontiers of theoretical physics. While its significance in quantum information science and quantum computing is well established \cite{RN196}, 
recent decades have witnessed remarkable extensions of its applicability. This measure now provides critical insights, ranging from black hole thermodynamics \cite{RN116,RN119,RN115} and holographic principles \cite{RN49} to condensed matter systems \cite{RN99,RN100} and 
quantum field theory itself \cite{RN195,RN112,RN117,RN80}.

Traditional studies of entanglement entropy have predominantly focused on real-space bipartitions, motivated by the local nature of quantum field theory (QFT)\cite{RN47,RN120,RN113}. However, a growing body of work has recently highlighted the profound implications
 of momentum-space entanglement entropy \cite{RN121,RN65,RN108,RN72,RN87,RN167,RN195}. This momentum-space perspective naturally connects with Wilsonian renormalization group concepts, since the reduced density matrix for low-energy modes directly corresponds to the Wilsonian 
 effective action,  {as first emphasized in Ref.~\cite{RN121}. }
 
  {In Ref.~\cite{RN121}, the authors propose two kinds of methods for the perturbative calculation of several types of entanglement measures and employ the time-indenpendent perturbation theory of ordinary quantum mechanics 
 to evaluate momentum space entanglement of $\phi^n$ model. They uncover an intriguing connection between renormalizability and the divergence properties of the mutual information, thereby providing key insights into the implications of momentum-space entanglement for renormalization group flows. 
 This approach was subsequently employed in Ref.~\cite{RN72} to analyze the momentum space entanglement property of interacting fermions at finite density in one spatial dimension. There, divergences in single-mode entanglement were identified as the momentum approached the Fermi surface, and several RG predictions were verified using mutual information.}
 
  {
 In Ref.~\cite{RN108} it was demonstrated that the Wilsonian formulation of the renormalization group defines a quantum channel
acting on the momentum-space density matrices of a quantum field theory. This result implies that the vacua of theories at a fixed
point contain no entanglement between momentum scales. Such findings naturally raise the question of to what extent specific RG flows are reflected in momentum-space entanglement entropy. Motivated by these developments, in this work we focus on momentum-space entanglement and explore systematic methods for calculating such entanglement measures.}

 {   Another seminal formulation proposed in Ref.~\cite{RN121} establishes a connection between RG and reduced density matrix through the explicit representation of the momentum-space reduced density matrix,}
\begin{equation}\label{eq:reduced-density-matrix}
  \bra{\varphi_{\vk}}\rho_{\mu}\ket{\tilde{\varphi}_{\vk}} = \lim_{\beta\to\infty}\frac{1}{Z(\beta)}\int_{\phi_{\vk}(0)=\varphi_{\vk}}^{\phi_{\vk}(\beta)=\tilde{\varphi}_{\vk}}\D\phi_{\vk}(\tau)\e^{-S_{\mu}^{\beta}}.
\end{equation}
 {However, this formalism was not explicitly used in Ref.~\cite{RN121}}, subsequent work by Ref.~\cite{RN65} developed a novel replica-trick approach for computing entanglement and Rényi entropies directly from the effective action based on this representation, revealing intriguing diagrammatic interpretations that challenge conventional perturbative understanding.
Although this method was originally developed for bosonic fields and successfully reproduces results obtained through conventional density matrix diagonalization in \cite{RN121}, its extension to fermionic systems  which is the topic of this paper, requires careful reconsideration. The Grassmannian nature of fermionic
 fields introduces significant complexities, rendering the path integral representation of the reduced density matrix substantially more intricate than in bosonic cases. 

While momentum-space entanglement in fermionic systems has been extensively studied in condensed matter physics \cite{RN86,RRN72,RN197,RN198,RN199,RN200,RN201},our primary interest lies in relativistic quantum field theories.
In this work, we extend the framework in Ref.~\cite{RN65} to fermionic systems, investigating whether analogous Feynman diagrammatic structures emerge in the computation of momentum-space entanglement entropy. Our analysis demonstrates affirmative results with unexpected new features. 


The remainder of this paper is organized as follows: Section~\ref{sec-2} presents a generalized formulation adapted for fermionic path integrals. Section~\ref{sec-3} applies this formalism to a four-fermion interaction model previously studied in Ref.~\cite{RN121}, while Section~\ref{sec-4} 
develops modified Feynman rules that capture the distinctive characteristics of fermionic entanglement entropy. Notably, we identify novel diagrammatic contributions whose physical origin requires further investigation. Technical details of the Grassmannian functional calculations are provided in Appendix~\ref{appendix}.

\section{Density matrix and replica trick for fermions}\label{sec-2}

The decomposition of physical degrees of freedom into high- and low-energy modes forms a cornerstone of the Wilsonian renormalization framework \cite{RN66,RN67}. Within this paradigm, systematically integrating out high-momentum degrees of freedom generates renormalization group (RG) flows for the coupling constants that govern low-energy effective interactions. This fundamental connection motivates the formal representation of reduced density matrices for low-momentum modes [Eq.~\ref{eq:reduced-density-matrix}], from which entanglement entropies can be systematically computed. In this work, we extend this formalism to fermionic systems.

Conventional Wilsonian RG implementations typically employ Wick-rotated Euclidean four-momentum integrals to derive manifestly Lorentz-invariant effective actions. By contrast, our density-matrix formulation requires a different approach. Following Ref.~\cite{RN121}, we construct the vacuum state using a Euclidean path integral in which spatial high-momentum modes are integrated out. Crucially, this procedure generates nonlocal contributions to the effective action—features shown to be essential for encoding entanglement signatures in bosonic systems Ref.~\cite{RN65}. As we demonstrate below, this key characteristic also persists in fermionic systems.

\subsection{Reduced density matrix for low-momentum degrees of freedom for fermions}

The path-integral representation of the vacuum state has been widely employed to compute Rényi entropies via the replica trick \cite{RN120}. 
This construction entends naturally to fermionic systems, though with crucial distinctions arising from the Grassmannian nature of the path integrals~\cite{RN112}. 

To set the stage, consider a bare action $S_{\Lambda}$ of a Dirac fermion field $\hat{\psi}(\vec{x})$ whose Fourier components are $\hat{\psi}_{\vk}$. We impose an ultraviolet cutoff $\Lambda$ on the spatial momentum such that $|\vk|<\Lambda$ {(In this paper bold letter $\vk$ denote spatial momentum).}
The ground state $\ket{\Psi_{\Lambda}^0}$ can be represented by the density matrix $\rho_{\Lambda}^0=\ket{\Psi_{\Lambda}^0}\bra{\Psi_{\Lambda}^0}$. 
Choosing a basis formed by eigenvectors of the field operator, $\hat{\psi}_{\vk}\ket{\psi_{\vk}}=\psi_{\vk}\ket{\psi_{\vk}}$, where $\psi_{\vk}$ is any Grassmann valued
function. Then the matrix element of the vacuum density operator in the momentum space are given by Ref.~\cite{RN112},

\begin{align}\label{eq-full-density-matrix}
  &\phantom{\mathrel{=}} \bra{\theta_{\vk}}\rho_{\Lambda}\ket{\tilde{\theta}_{\vk}} \nonumber\\
  & = \lim_{\beta\to\infty}\frac{1}{Z(\beta)}\int_{\psi_{\vk}(0)=\theta_{\vk}}^{\psi_{\vk}(\beta)=\tilde{\theta}_{\vk}}\D\bar{\psi}_{\vk}(\tau)\D\psi_{\vk}(\tau)\e^{-S_E(\psi_{\vk},\bar{\psi}_{\vk})}.
\end{align}
where $S_\E$ is the Euclidean action and $Z(\beta)$ is the partition function that normalizas the path integral.

 {It was first pointed out in \cite{RN121} that the density matrix $\rho_{\mu}$ for the degrees of freedom with momentum $\vk<\mu$ is naturally associated with the Wilsonian effective action $S_{\mu}$(see eq.\ref{eq:reduced-density-matrix}). The latter is obtained from the bare
action $S_{\Lambda}$ by integrating out all field modes
with momentum $\vk $ such that $ \vk \ge \mu$, where $\mu$ defines a momentum scale in the range $[0, \Lambda]$ that seprate the momentum modes into two disjoint subsets.} Accordingly, we decompose the fields in momentum space into low- and high-momentum modes,
\begin{equation}
  \psi_{\vk} = \psi_{0,\vk} + \psi_{1,\vk},
\end{equation}
with
\begin{equation}
	\begin{cases}
		\psi_{0,\vk}=\psi_{\vk} & \text{if $|\vk|\le\mu$, otherwise 0}, \\
		\psi_{1,\vk}=\psi_{\vk} & 	\text{if $|\vk|>\mu$, otherwise 0}.
	\end{cases}
\end{equation}
and similarly for $\bar{\psi}_{\vk}$. Then the reduced density matrix of low-momentum modes is obtained by trace out high-momentum modes in eq. \ref{eq-full-density-matrix},
\begin{widetext}
\begin{align}
  &\phantom{\mathrel{=}} \bra{\theta_{0, \vk}}\rho_{\mu}\ket{\tilde{\theta}_{0,\vk}} = \lim_{\beta\to\infty}\frac{1}{Z(\beta)}\int_{\psi_{0,\vk}(0)=\theta_{0,\vk}}^{\psi_{0,\vk}(\beta)=\tilde{\theta}_{0,\vk}}\D\bar{\psi}_{0,\vk}(\tau)\D\psi_{0,\vk}(\tau)\e^{-S_{\mu}(\psi_{0, \vk},\bar{\psi}_{0, \vk})},
\end{align}
\end{widetext}
where the low energy effective action at scale $\mu$ is defined by,
\begin{equation}
  \e^{-S_{\mu}(\psi_{0, \vk},\bar{\psi}_{0, \vk})} = \int_{a.p.}\D\bar{\psi}_{1,\vk}\D\psi_{1,\vk}\e^{-S_E(\psi_{\vk},\bar{\psi}_{\vk})}.
\end{equation}
Here the subscript "a.p." indicates that the high-momentum modes of the fermion field satisfy antiperiodic boundary conditions, as is standard in the evaluation of fermionic thermal partition functions.
This representation allows the low-energy effective action of a fermionic system to be calculated perturbatively.


In contrast to the conventional approach of Ref.~\cite{RN112}, where antiperiodicity is imposed on the low-momentum sector, we instead assign the minus sign arising from the Grassmann trace to the high-momentum modes. This choice enforces antiperiodic boundary conditions exclusively on the high-momentum fields. Consequently, the path integral is restricted to antiperiodic configurations of the high-momentum modes. After performing this momentum-shell integration, the high-momentum degrees of freedom are systematically eliminated, leaving a reduced system in which only the low-momentum modes remain as fundamental dynamical variables. This construction thereby establishes an effective theory for the infrared regime.

 {It is worth emphasizing that the low-energy effective action is a highly valuable object. On the one hand, its dependence on the scale 
$\mu$ provides key insights into the insensitivity of low-energy physics to the details of the ultraviolet description. On the other hand, it captures the complete information about the entanglement between high- and low-momentum modes. Thus, the Wilsonian effective action serves as a crucial bridge between renormalization and momentum-space entanglement—a connection that is intriguing yet far from fully understood. } 

\subsection{Entanglement measures from the effective action}


Having established the low-energy effective action, we now proceed to compute the Rényi entropy using the replica technique, which fundamentally reduces to evaluating $\text{Tr} \rho_{\mu}^n$. Our methodology follows 
the prescription outlined in Ref.~\cite{RN65}, extended here to accommodate fermionic systems. The key step involves constructing a modified partition function, defined as { (see Ref.~\cite{RN65} for a detailed description)}:
\begin{equation}
  Z_n(\mu,\beta) \equiv \int_{n\beta}\D\bar{\psi}_{0,\vk}(\tau)\D\psi_{0.\vk}(\tau)\e^{-\sum_{u=0}^{n-1}\int_{u\beta}^{(u+1)\beta}\d\tau L_{\mu}^{\beta}},
\end{equation}
where the effective Lagrangian density $L_{\mu}^{\beta}$ integrates to the action $S_{\mu}^{\beta} = \int_{0}^{\beta} \mathrm{d}\tau L_{\mu}^{\beta}$. This construction arises from a two-step geometric manipulation of the Euclidean time domain:
first, shifting the Euclidean time range for each replica of the reduced density matrix; and second,  cyclically gluing these $n$ copies across adjacent temporal boundaries. Crucially, for fermionic systems, the anti-periodicity condition must now be imposed 
on the low-momentum modes over the extended interval $[0, n\beta]$, in contrast to conventional bosonic treatments. With this framework established, we can write:

\begin{equation}\label{eq-power-reduced-density-matrix}
  \text{Tr}\rho_{\mu}^n = \lim_{\beta\to\infty}\frac{Z_n(\mu,\beta)}{[Z(\mu,\beta)]^n},
\end{equation}
where the partition function is redefined by discarding terms which is at zeroth order in the low-energy effective action,
\begin{equation}
  Z(\mu, \beta) \equiv \int_{\beta}\D\bar{\psi}_{0,\vk}(\tau)\D\psi_{0,\vk}(\tau)\e^{-S_{\mu}^{\beta}(\psi_{0,\vk},\bar{\psi}_{0,\vk})}.
\end{equation}

The central object of interest is the “glued” effective action, defined as
 $$S_{\mu,r}^{\beta}\equiv\sum_{u=0}^{n-1}\int_{u\beta}^{(u+1)\beta}\d\tau L_{\mu}^{\beta}.$$

  At second order, this glued action decomposes into two contributions:one corresponding to the original effective action evaluated at temperature $1/n\beta$, and the other a nonlocal component that contributes directly to the Rényi entropy.
Accordingly, we may write
\begin{equation}\label{eq-decomposition-of-effective-action}
  S_{\mu,r}^{\beta}\equiv S_{\mu}^{n\beta} + S_{\mu,\text{nl}}^{n\beta}.
\end{equation}

Furthermore, to facilitate a perturbative expansion, we define the unperturbed partition function of the low-momentum modes as
\begin{equation}
  Z_0(\mu,\beta)\equiv\int_{\beta}\D\bar{\psi}_{0,\vk}(\tau)\psi_{0,\vk}(\tau)\e^{S_{\mu}^{\beta,(0)}(\bar{\psi}_{0,\vk},\psi_{0,\vk})},
\end{equation}
where $S_{\mu}^{\beta,(0)}$ represent the free part of the effective action.

Then we can write Eq.~\ref{eq-power-reduced-density-matrix} as:
\begin{align}
  \text{Tr}\rho_{\mu}^n & = \lim_{\beta\to\infty}\frac{Z_n(\mu,\beta)}{[Z(\mu,\beta)]^n} \nonumber \\
  & = \lim_{\beta\to\infty}\frac{Z_n(\mu,\beta)}{Z(\mu,n\beta)} \nonumber \\
  & = \lim_{\beta\to\infty}\frac{Z_n(\mu,\beta)/Z_0(\mu,n\beta)}{Z(\mu,n\beta)/Z_0(\mu,n\beta)}. 
\end{align}
Finally he Rényi entropies can be expressed as
\begin{align}\label{eq-Renyi-entropy}
  H_n(\mu) & = \frac{1}{1-n}\log\,\tr\,\rho_\mu^n \nonumber \\
  & = \frac{1}{n-1}\lim_{\beta\to\infty}(\mathrm{log}\,\frac{Z(\mu,n\beta)}{Z_0(\mu,n\beta)}-\mathrm{log} \,\frac{Z_n(\mu,\beta)}{Z_0(\mu,n\beta)}).
\end{align}

Using the decomposition of Eq.~\ref{eq-decomposition-of-effective-action}, the perturbative expansion to leading order yields
\begin{equation}\label{eq-Renyi-entropy-perturbative}
	H_n(\mu)=\frac{1}{n-1}\lim_{\beta\to\infty}\braket{S_{\mu,\text{nl}}^{n\beta}}.
\end{equation}
In Ref.~\cite{RN65},the authors established a general theorem demonstrating that if the effective action consists solely of local terms in Euclidean time, the Rényi entropies vanish identically at the nonperturbative level. Our findings extend this conclusion by showing that even when the effective action contains both local and nonlocal temporal contributions, the entanglement structure is still entirely governed by the nonlocal sector within perturbation theory. Specifically, nonlocal contributions persist as the exclusive perturbative source of quantum entanglement, while local terms alone fail to generate nontrivial entropy corrections. This distinction highlights the intrinsic link between temporal nonlocality in the effective action and the emergence of quantum correlations quantified by Rényi entropies.

In practice, exact expressions for the effective action are generally inaccessible. As emphasized in Ref.~\cite{RN65}, the appearance of nontrivial entanglement between distinct momentum modes requires identifying the lowest-order effective action that incorporates nonlocal temporal terms. In our case, such contributions first arise at the second-order expansion of the effective action.

A subtle technical point arises in deriving Eq.~\ref{eq-Renyi-entropy-perturbative}: the perturbative procedure is applied at two distinct stages — first, in the derivation of the effective action itself, and second, in the evaluation of the composite path integral for the replica system. This dual perturbative treatment underscores the importance of consistently truncating approximations while retaining entanglement-sensitive terms across both computational steps.

\section{Application to $(\bar{\psi}\psi)^2$ theory}\label{sec-3}
In this section, we apply the preceding formalism to a four-fermion interaction model. While detailed derivations are provided in Appendix \ref{appendix}, we summarize the key results at each computational stage below.

{This interaction model has been widely studied in various contexts. For instance, Ref.~\cite{RN121} computed the entanglement entropy via time-independent perturbation theory, yielding leading-order results. The same framework was later extended to finite-density fermionic systems in Ref.~\cite{RRN72}, but such approaches rely on explicit density-matrix diagonalization and are not suitable for higher-order corrections. A related study, Ref.~\cite{RN86}, employed the replica trick to derive a systematic expansion of the Rényi entropies $S_n(A)$ for momentum-space subsets near the Fermi surface. Although conceptually similar to our approach, their diagrammatic representation (“pants diagrams”) is not analytically tractable, whereas our method provides a calculable framework.}

We consider the Euclidean action,
\begin{widetext}
  \begin{equation}
    S_{\Lambda}^{\beta}[\psi_{\vk}] = \sum_{K,Q}\bar{\psi}_{K}D_{K,Q}\Psi_Q-g\frac{\beta}{V}\sum_{K,K',Q,Q'}\bar{\psi}_K\psi_Q\bar{\psi}_{K'}\psi_{Q'}\delta_{K+K'-Q-Q',0},
  \end{equation}
\end{widetext}
with Euclidean momentum $K^{\mu}\equiv(-\i\omega_j, \vk)$ where $\omega_j=(2j+1)\pi/\beta$ are the fermion Matsubara frenquencies. The spacetime dimension $d$ is left arbitrary. This Euclidean action can be derived from the Lagrangian,
\begin{equation}
  \mathcal{L} = \bar{\psi}(\i\slashed{\partial}-m)\psi+g(\bar{\psi}\psi)^2,
\end{equation}
by transforming the fields to momentum space via the standard Fourier transforms in finite-temperature field theory:
\begin{equation}
	\begin{aligned}
		\psi(X)&=\frac{1}{\sqrt{V}}\sum_K\e^{-\i K\cdot X}\psi_K\\,
		\bar{\psi}(X)&=\frac{1}{\sqrt{V}}\sum_K\e^{\i K\cdot X}\bar{\psi}_K,
	\end{aligned}
\end{equation}
with Euclidean coordinate defined as $X^{\mu}\equiv (-\i\tau, \bm{x})$ then we find the Euclidean action above with $D_{K,Q}\equiv\frac{\S_0^{-1}(K)}{T}\delta_{K,Q}$, where
$\S_0^{-1}(K)\equiv m-\slashed{K}$ is the thermal fermion propagator.

Given the Euclidean action, the first step is to calculate the low energy effective action. To order $g^2$, there are four Feynman diagramns as in Figs.~\multiref{fig:order1fermion}{fig:3oforder2psi4} that contribute, 
\begin{figure}[!htbp]
    \centering
    \begin{center}
      \begin{tikzpicture}
        \begin{feynman}
          \vertex (a);
          \vertex [above right=of a] (b);
          \vertex [below right=of b] (c);
          \vertex [above=of b] (d);
          \diagram*{
          (a) -- [fermion, momentum={[arrow shorten = 0.3]$K_1$}] (b),
          (b) -- [fermion, momentum={[arrow shorten = 0.3]$K_2$}] (c),
          (b) -- [half left, charged scalar, momentum={[arrow shorten = 0.3]$K_3$}] (d),
          (d) -- [half left, charged scalar] (b),
          };
          \filldraw [black] (b) circle [radius=2pt];
        \end{feynman}
      \end{tikzpicture}
      \captionsetup{justification=justified, singlelinecheck=false}
      \caption{ {Order $g$ contribution to the low-energy effective action: momentum conservation prevents entanglement between high- and low-momentum modes, so this term is omitted.}}
      \label{fig:order1fermion}
    \end{center}
\end{figure}
\begin{figure}[!htbp]
  \centering
  \begin{tikzpicture}
      \begin{feynman}
          \vertex (a);
          \vertex [right=of a] (b);
          \vertex [right=of b] (c);
          \vertex [right=of c] (d);
          \diagram*{
          (a) -- [fermion, momentum={[arrow shorten = 0.3]$K_1$}] (b),
          (b) -- [half left, charged scalar, momentum={[arrow shorten = 0.3]$K_3$}] (c),
          (c) -- [half left, charged scalar, momentum={[arrow shorten = 0.3]$K_5$}] (b),
          (b) -- [charged scalar, momentum'={[arrow shorten = 0.3, arrow distance = 1mm]\footnotesize $K_4$}] (c),
          (c) -- [fermion, momentum={[arrow shorten = 0.3]$K_2$}] (d),
          };
          \filldraw [black] (b) circle [radius=2pt];
          \filldraw [black] (c) circle [radius=2pt];
      \end{feynman}
  \end{tikzpicture}
  \caption{ {Order $g^2$ two-point contribution to the low-energy effective action. In fermionic systems, two types of contractions of the legs are possible, as shown in Fig.~\ref{fig:contractiontype2}.} }
  \label{fig:1oforder2psi4}
\end{figure}
\begin{figure}
  \centering
  \begin{tikzpicture}
    \begin{feynman}
      \vertex (a);
      \vertex [below right=of a] (b);
      \vertex [below left=of b] (c);
      \vertex [right=of b] (d);
      \vertex [above right=of d] (e);
      \vertex [below right=of d] (f);
      \diagram*{
      (a) --[fermion, momentum={[arrow shorten = 0.3, arrow distance = 2mm]$K_1$}] (b),
      (b) --[fermion, momentum={[arrow shorten = 0.3, arrow distance = 2mm]$K_2$}] (c),
      (b) -- [half left, charged scalar, momentum={[arrow shorten = 0.35, arrow distance = 2mm] \footnotesize $K_5$}] (d),
      (d) -- [half left, charged scalar, momentum={[arrow shorten = 0.35, arrow distance = 2mm] \footnotesize $K_6$}] (b),
      (d) -- [anti fermion, rmomentum={[arrow shorten = 0.3, arrow distance = 2mm]$K_3$}] (e),
      (d) -- [fermion, momentum'={[arrow shorten = 0.3, arrow distance = 2mm]$K_4$}] (f),
      };
      \filldraw [black] (b) circle [radius=2pt];
      \filldraw [black] (d) circle [radius=2pt];
    \end{feynman}
  \end{tikzpicture}
  \caption{ {Order $g^2$ four-point contribution to the low-energy effective action, with three possible fermionic leg contractions (see Fig.~\ref{fig:contraction4type3}).} }
  \label{fig:2oforder2psi4}
\end{figure}
\begin{figure}[!htbp]
  \centering
  \begin{tikzpicture}
    \begin{feynman}
      \vertex (a);
      \vertex [below right=of a] (b);
      \vertex [left=of b] (c);
      \vertex [below left=of b] (d);
      \vertex [right=of b] (e);
      \vertex [above right=of e] (f);
      \vertex [right=of e] (g);
      \vertex [below right=of e] (h);
      \diagram*{
      (a) --[fermion, momentum={[arrow shorten = 0.3, arrow distance = 1.5mm] \footnotesize$K_1$}] (b),
      (c) --[fermion, momentum={[arrow shorten = 0.3, arrow distance = 1.5mm] \footnotesize$K_2$}] (b),
      (d) --[anti fermion, rmomentum'={[arrow shorten = 0.3, arrow distance = 1.5mm] \footnotesize$K_3$}] (b),
      (b) --[charged scalar, momentum={[arrow shorten = 0.3, arrow distance = 1.5mm] \footnotesize$K_7$}] (e),
      (f) --[fermion, momentum'={[arrow shorten = 0.3, arrow distance = 1.5mm] \footnotesize$K_4$}] (e),
      (g) --[anti fermion, rmomentum'={[arrow shorten = 0.3, arrow distance = 1.5mm] \footnotesize$K_5$}] (e),
      (h) --[anti fermion, rmomentum={[arrow shorten = 0.3, arrow distance = 1.5mm] \footnotesize$K_6$}] (e),
      };
      \filldraw [black] (b) circle [radius=2pt];
      \filldraw [black] (e) circle [radius=2pt];
    \end{feynman}
  \end{tikzpicture}
  \caption{ {Order $g^2$ six-point contribution to the low-energy effective action, denoted by $S_{\mu}^{\beta, (6)}$.}}
  \label{fig:3oforder2psi4}
\end{figure}

Note that due to the Dirac indices, there are several types of contractions for the two-point and four-point effective actions (see Appendix~\ref{appendix}). In particular, the two-point diagram can be contracted as illustrated in Fig.~\ref{fig:contractiontype2}.
\begin{figure*}
	\centering
	\begin{tikzpicture}
		\begin{feynman}
			\vertex (a);
			\vertex [right=of a] (b);
			\vertex [above right=of b] (c);
			\vertex [below right=of b] (d);
			\vertex [right=of b] (e);
			\vertex [right=of e] (f);
			\vertex [right=of f] (g);
			\vertex [right=of g] (h);
			\vertex [above left = of g] (i);
			\vertex [below left = of g] (j);
			\diagram*{
			(a) --[fermion, momentum={[arrow shorten = 0.3, arrow distance = 1.5mm] \footnotesize$\alpha$}] (b),
			(b) --[fermion, momentum={[arrow shorten = 0.3, arrow distance = 1.5mm] \footnotesize$\beta$}] (c),
			(b) --[fermion, momentum={[arrow shorten = 0.3, arrow distance = 1.5mm] \footnotesize$\alpha$}] (e),
			(d) --[fermion, momentum={[arrow shorten = 0.3, arrow distance = 1.5mm] \footnotesize$\beta$}] (b),
			(f) --[fermion, momentum={[arrow shorten = 0.3, arrow distance = 1.5mm] \footnotesize$\sigma$}] (g),
			(i) --[fermion, momentum={[arrow shorten = 0.3, arrow distance = 1.5mm] \footnotesize$\rho$}] (g),
			(g) --[fermion, momentum={[arrow shorten = 0.3, arrow distance = 1.5mm] \footnotesize$\rho$}] (j),
			(g) --[fermion, momentum={[arrow shorten = 0.3, arrow distance = 1.5mm] \footnotesize$\sigma$}] (h),
			(c) --[scalar, out = 45, in = 135] (i),
			(e) --[scalar] (f),
			(d) --[scalar, out = -45, in = -135] (j)
			};
			\filldraw [black] (b) circle [radius=2pt];
			\filldraw [black] (g) circle [radius=2pt];
		\end{feynman}
	\end{tikzpicture}
  \hspace{1cm}
	\begin{tikzpicture}
		\begin{feynman}
			\vertex (a);
			\vertex [right=of a] (b);
			\vertex [above right=of b] (c);
			\vertex [below right=of b] (d);
			\vertex [right=of b] (e);
			\vertex [right=of e] (f);
			\vertex [right=of f] (g);
			\vertex [right=of g] (h);
			\vertex [above left = of g] (i);
			\vertex [below left = of g] (j);
			\diagram*{
			(a) --[fermion, momentum={[arrow shorten = 0.3, arrow distance = 1.5mm] \footnotesize$\alpha$}] (b),
			(b) --[fermion, momentum={[arrow shorten = 0.3, arrow distance = 1.5mm] \footnotesize$\beta$}] (c),
			(b) --[fermion, momentum={[arrow shorten = 0.3, arrow distance = 1.5mm] \footnotesize$\alpha$}] (e),
			(d) --[fermion, momentum={[arrow shorten = 0.3, arrow distance = 1.5mm] \footnotesize$\beta$}] (b),
			(f) --[fermion, momentum={[arrow shorten = 0.3, arrow distance = 1.5mm] \footnotesize$\sigma$}] (g),
			(i) --[fermion, momentum={[arrow shorten = 0.3, arrow distance = 1.5mm] \footnotesize$\rho$}] (g),
			(g) --[fermion, momentum={[arrow shorten = 0.3, arrow distance = 1.5mm] \footnotesize$\rho$}] (j),
			(g) --[fermion, momentum={[arrow shorten = 0.3, arrow distance = 1.5mm] \footnotesize$\sigma$}] (h),
			(c) --[scalar, out = 45, in = 135] (f),
			(e) --[scalar, out = 45, in = 135] (i),
			(d) --[scalar, out = -45, in = -135] (j)
			};
			\filldraw [black] (b) circle [radius=2pt];
			\filldraw [black] (g) circle [radius=2pt];
		\end{feynman}
	\end{tikzpicture}
	\caption{ {Two contractions of the order $g^2$ two-point contribution to the low-energy effective action, distinguished by Dirac indices and denoted $S_{\mu}^{\beta, (2, 1)}$ and $S_{\mu}^{\beta, (2, 2)}$ respectively.}}
	\label{fig:contractiontype2}
\end{figure*}

And the 4 point diagram has three types of contractions as in Fig.~\ref{fig:contraction4type3}.

\begin{figure*}
    \centering
    \begin{tikzpicture}
      \begin{feynman}
        \vertex (a);
        \vertex [above right = of a] (b);
        \vertex [above left = of a] (d);
        \vertex [below right = of a] (c);
        \vertex [below left = of a] (e);
        \vertex [right = of a] (k);
        \vertex [right = of k] (f);
        \vertex [above right = of f] (i);
        \vertex [below right=of f] (j);
        \vertex [above left = of f] (g);
        \vertex [below left = of f] (h);
        \diagram*{
        (d) --[fermion, momentum={[arrow shorten = 0.3, arrow distance = 1.5mm] \footnotesize$\beta$}] (a),
        (a) --[fermion, momentum={[arrow shorten = 0.3, arrow distance = 1.5mm] \footnotesize$\alpha$}] (b),
        (c) --[fermion, momentum={[arrow shorten = 0.3, arrow distance = 1.5mm] \footnotesize$\alpha$}] (a),
        (a) --[fermion, momentum={[arrow shorten = 0.3, arrow distance = 1.5mm] \footnotesize$\beta$}] (e),
        (g) --[fermion, momentum={[arrow shorten = 0.3, arrow distance = 1.5mm] \footnotesize$\rho$}] (f),
        (f) --[fermion, momentum={[arrow shorten = 0.3, arrow distance = 1.5mm] \footnotesize$\rho$}] (h),
        (i) --[fermion, momentum={[arrow shorten = 0.3, arrow distance = 1.5mm] \footnotesize$\sigma$}] (f),
        (f) --[fermion, momentum={[arrow shorten = 0.3, arrow distance = 1.5mm] \footnotesize$\sigma$}] (j),
        (b) --[scalar, out = 45, in = 135] (g),
        (h) --[scalar, out = -135, in = -45] (c)
        };
        \filldraw [black] (a) circle [radius=2pt];
        \filldraw [black] (f) circle [radius=2pt];
      \end{feynman}
    \end{tikzpicture}
    \hspace{.4cm}
    \begin{tikzpicture}
      \begin{feynman}
        \vertex (a);
        \vertex [above right = of a] (b);
        \vertex [above left = of a] (d);
        \vertex [below right = of a] (c);
        \vertex [below left = of a] (e);
        \vertex [right = of a] (k);
        \vertex [right = of k] (f);
        \vertex [above right = of f] (i);
        \vertex [below right=of f] (j);
        \vertex [above left = of f] (g);
        \vertex [below left = of f] (h);
        \diagram*{
        (d) --[fermion, momentum={[arrow shorten = 0.3, arrow distance = 1.5mm] \footnotesize$\beta$}] (a),
        (a) --[fermion, momentum={[arrow shorten = 0.3, arrow distance = 1.5mm] \footnotesize$\alpha$}] (b),
        (c) --[fermion, momentum={[arrow shorten = 0.3, arrow distance = 1.5mm] \footnotesize$\alpha$}] (a),
        (a) --[fermion, momentum={[arrow shorten = 0.3, arrow distance = 1.5mm] \footnotesize$\beta$}] (e),
        (g) --[fermion, momentum={[arrow shorten = 0.3, arrow distance = 1.5mm] \footnotesize$\sigma$}] (f),
        (f) --[fermion, momentum={[arrow shorten = 0.3, arrow distance = 1.5mm] \footnotesize$\rho$}] (h),
        (i) --[fermion, momentum={[arrow shorten = 0.3, arrow distance = 1.5mm] \footnotesize$\rho$}] (f),
        (f) --[fermion, momentum={[arrow shorten = 0.3, arrow distance = 1.5mm] \footnotesize$\sigma$}] (j),
        (b) --[scalar, out = 45, in = 135] (g),
        (h) --[scalar, out = -135, in = -45] (c)
        };
        \filldraw [black] (a) circle [radius=2pt];
        \filldraw [black] (f) circle [radius=2pt];
      \end{feynman}
    \end{tikzpicture}
    \hspace{.4cm}
    \begin{tikzpicture}
      \begin{feynman}
        \vertex (a);
        \vertex [above right = of a] (b);
        \vertex [above left = of a] (d);
        \vertex [below right = of a] (c);
        \vertex [below left = of a] (e);
        \vertex [right = of a] (k);
        \vertex [right = of k] (f);
        \vertex [above right = of f] (i);
        \vertex [below right=of f] (j);
        \vertex [above left = of f] (g);
        \vertex [below left = of f] (h);
        \diagram*{
        (d) --[fermion, momentum={[arrow shorten = 0.3, arrow distance = 1.5mm] \footnotesize$\beta$}] (a),
        (a) --[fermion, momentum={[arrow shorten = 0.3, arrow distance = 1.5mm] \footnotesize$\beta$}] (b),
        (c) --[fermion, momentum={[arrow shorten = 0.3, arrow distance = 1.5mm] \footnotesize$\alpha$}] (a),
        (a) --[fermion, momentum={[arrow shorten = 0.3, arrow distance = 1.5mm] \footnotesize$\alpha$}] (e),
        (g) --[fermion, momentum={[arrow shorten = 0.3, arrow distance = 1.5mm] \footnotesize$\sigma$}] (f),
        (f) --[fermion, momentum={[arrow shorten = 0.3, arrow distance = 1.5mm] \footnotesize$\rho$}] (h),
        (i) --[fermion, momentum={[arrow shorten = 0.3, arrow distance = 1.5mm] \footnotesize$\rho$}] (f),
        (f) --[fermion, momentum={[arrow shorten = 0.3, arrow distance = 1.5mm] \footnotesize$\sigma$}] (j),
        (b) --[scalar, out = 45, in = 135] (g),
        (h) --[scalar, out = -135, in = -45] (c)
        };
        \filldraw [black] (a) circle [radius=2pt];
        \filldraw [black] (f) circle [radius=2pt];
      \end{feynman}
    \end{tikzpicture}
  \caption{ {Three contractions of the order $g^2$ four-point contribution to the low-energy effective action, distinguished by Dirac indices and denoted $S_{\mu}^{\beta, (4, 1)}$, $S_{\mu}^{\beta, (4, 2)}$ and $S_{\mu}^{\beta, (4, 3)}$ respectively.}} 
  \label{fig:contraction4type3}
\end{figure*}        
 
From these diagram we can write down the effective action as
\begin{align}
	S_{\mu}^{\beta} & = S_{0, \E}^{\beta}+S_{\mu}^{\beta,(2,1)} + S_{\mu}^{\beta,(2,2)} \nonumber\\
  & \quad\quad + S_{\mu}^{\beta,(4,1)}+S_{\mu}^{\beta,(4,2)}+S_{\mu}^{\beta,(4,3)}+S_{\mu}^{\beta,(6)},
\end{align}
and the "glued" effective action as
\begin{align}
  S_{\mu, \r}^{\beta} & = S_{0, \E}^{n\beta}+S_{\mu,\r}^{\beta,(2,1)}+S_{\mu,\r}^{\beta,(2,2)} \nonumber \\
  &\quad\quad + S_{\mu,\r}^{\beta,(4,1)}+S_{\mu,\r}^{\beta,(4,2)}+S_{\mu,\r}^{\beta,(4,3)}+S_{\mu,\r}^{\beta,(6)},
\end{align}
see appendix.~\ref{appendix} for explicit expressions. 

Then according to Eq.~\ref{eq-Renyi-entropy}, we can write the Rényi entropy as Eq.~\ref{eq-renyi-4-fermion}. 
\begin{widetext}
  \begin{equation}\label{eq-renyi-4-fermion}
    \begin{aligned}
      \mathrel{\phantom{=}} H_n(\mu) & = \frac{1}{n-1}\lim_{\beta\to\infty}\left(\braket{S_{\mu,r}^{n\beta,(2,1)}}+\braket{S_{\mu,r}^{n\beta,(2,2)}}+\braket{S_{\mu,r}^{n\beta,(4,1)}}+\braket{S_{\mu,r}^{n\beta,(4,2)}}+\braket{S_{\mu,r}^{n\beta,(4,3)}}+\braket{S_{\mu,r}^{n\beta,(6)}}\right. \\ 
      &\mathrel{\phantom{=}}\; \left.-\braket{S_{\mu}^{n\beta,(2,1)}}-\braket{S_{\mu}^{n\beta,(2,2)}}-\braket{S_{\mu}^{n\beta,(4,1)}}-\braket{S_{\mu}^{n\beta,(4,2)}}-\braket{S_{\mu}^{n\beta,(4,3)}}-\braket{S_{\mu}^{n\beta,(6)}}\right).
    \end{aligned}
  \end{equation}
\end{widetext}

As analytically shown in Appendix~\ref{appendix} and anticipated in Eq.~\ref{eq-decomposition-of-effective-action}, the composite ("glued") effective action admits a systematic decomposition as the sum of the original effective action at temperature $n\beta$ and an additional nonlocal term,This property holds for each individual Feynman diagram and establishes that nonlocal interactions are not merely necessary for generating entanglement, but constitute the sole source of momentum-mode entanglement.

The evaluation of the expectation values in Eq.~\ref{eq-renyi-4-fermion} corresponds to contracting the remaining low-momentum external legs in Figs.~\ref{fig:order1fermion}--\ref{fig:3oforder2psi4}, following a procedure analogous to the contraction scheme described in Ref.~\cite{RN65}. 
It should be emphasized, however, that these diagrams serve as schematic representations; the actual computations involve the full complexity of fermionic propagators.
\begin{figure}[!htbp]
  \begin{tikzpicture}[baseline=(a.base)]
      \begin{feynman}[every plain=very thick, every scalar=very thick]
          \vertex (a);
          \vertex [right=of a] (b);
          \vertex [right=of b] (c);
          \vertex [right=of c] (d);
          \diagram*{
          (a) -- [plain] (b),
          (b) -- [half left, scalar] (c),
          (c) -- [half left, scalar] (b),
          (b) -- [scalar] (c),
          (c) -- [plain] (d),
          };
          \filldraw [black] (b) circle [radius=2pt];
          \filldraw [black] (c) circle [radius=2pt];
      \end{feynman}
  \end{tikzpicture}

  \centering{\qquad $\Downarrow$ \qquad}

  \begin{tikzpicture}[baseline=(a.base)]
      \begin{feynman}[every scalar={very thick}, every plain={very thick}]
          \vertex (a);
          \vertex [right=of a] (b);
          \diagram*{
          (a) -- [half left, looseness=1.7, plain] (b);
          (b) -- [half left, looseness=1.7, scalar] (a);
          (a) -- [half left, looseness=0.8, scalar] (b);
          (b) -- [half left, looseness=0.8, scalar] (a)
          };
          \filldraw [black] (a) circle [radius=2pt];
          \filldraw [black] (b) circle [radius=2pt];
      \end{feynman}
  \end{tikzpicture}
  \caption{Schematic contractions of the two-point effective action}
\end{figure} 
\begin{figure}[!htbp]
  \begin{tikzpicture}[baseline=(b.base), scale=0.8]
      \begin{feynman}[every scalar=very thick, every plain=very thick]
          \vertex (a);
          \vertex [below right=of a] (b);
          \vertex [below left=of b] (c);
          \vertex [right=of b] (d);
          \vertex [above right=of d] (e);
          \vertex [below right=of d] (f);
          \diagram*{
          (a) --[plain] (b),
          (b) --[plain] (c),
          (b) -- [half left, scalar] (d),
          (d) -- [half left, scalar] (b),
          (d) -- [plain] (e),
          (d) -- [plain] (f),
          };
          \filldraw [black] (d) circle [radius=2pt];
          \filldraw [black] (b) circle [radius=2pt];
      \end{feynman}
  \end{tikzpicture}
            
  \qquad $\Downarrow$ \qquad
            
  \begin{tikzpicture}[baseline=(a.base)]
      \begin{feynman}[every scalar=very thick, every plain=very thick]
          \vertex (a);
          \vertex [right=of a] (b);
          \diagram*{
          (a) -- [half left, looseness=1.7, plain] (b),
          (b) -- [half left, looseness=1.7, plain] (a),
          (a) -- [half left, looseness=0.8, scalar] (b),
          (b) -- [half left, looseness=0.8, scalar] (a),
          };
          \filldraw [black] (a) circle [radius=2pt];
          \filldraw [black] (b) circle [radius=2pt];
      \end{feynman}
  \end{tikzpicture}
  \qquad +\qquad
  \begin{tikzpicture}[baseline=(a.base)]
      \begin{feynman}[every scalar=very thick, every plain=very thick]
          \vertex (a);
          \vertex [right=of a] (b);
          \vertex [right=of b] (c);
          \vertex [right=of c] (d);
          \diagram*{
          (a) -- [half left, plain] (b),
          (b) -- [half left, plain] (a),
          (c) -- [half left, plain] (d),
          (d) -- [half left, plain] (c),
          (b) -- [half left, scalar] (c),
          (c) -- [half left, scalar] (b),
          };
          \filldraw [black] (c) circle [radius=2pt];
          \filldraw [black] (b) circle [radius=2pt];
      \end{feynman}
  \end{tikzpicture}
  \caption{Schematic contractions of the four-point effective action}
  \label{fig:contraction-4-point-effective-action}
\end{figure}

\begin{figure}[!htbp]
  \begin{tikzpicture}[baseline=(b.base)]
    \begin{feynman}[every scalar=very thick, every plain=very thick]
        \vertex (a);
        \vertex [below right=of a] (b);
        \vertex [left=of b] (c);
        \vertex [below left=of b] (d);
        \vertex [right=of b] (e);
        \vertex [above right=of e] (f);
        \vertex [right=of e] (g);
        \vertex [below right=of e] (h);
        \diagram*{
        (a) --[plain] (b),
        (c) --[plain] (b),
        (d) --[plain] (b),
        (b) --[scalar] (e),
        (f) --[plain] (e),
        (g) --[plain] (e),
        (h) --[plain] (e),
        };
        \filldraw [black] (b) circle [radius=2pt];
        \filldraw [black] (e) circle [radius=2pt];
      \end{feynman}
  \end{tikzpicture}

  \qquad $\Downarrow$ \qquad

  \begin{tikzpicture}[baseline=(a.base)]
    \begin{feynman}[every scalar=very thick, every plain=very thick]
      \vertex (a);
      \vertex [right=of a] (b);
      \diagram*{
      (a) -- [half left, looseness=1.7, plain] (b),
      (b) -- [half left, looseness=1.7, scalar] (a),
      (a) -- [half left, looseness=0.8, plain] (b),
      (b) -- [half left, looseness=0.8, plain] (a),
    };
    \filldraw [black] (a) circle [radius=2pt];
    \filldraw [black] (b) circle [radius=2pt];
    \end{feynman}
  \end{tikzpicture}
  \qquad + \qquad
  \begin{tikzpicture}[baseline=(a.base)]
  \begin{feynman}[every scalar=very thick, every plain=very thick]
      \vertex (a);
      \vertex [right=of a] (b);
      \vertex [right=of b] (c);
      \vertex [right=of c] (d);
      \diagram*{
      (a) -- [half left, plain] (b),
      (b) -- [half left, plain] (a),
      (c) -- [half left, plain] (d),
      (d) -- [half left, plain] (c),
      (b) -- [half left, plain] (c),
      (c) -- [half left, scalar] (b),
      };
      \filldraw [black] (c) circle [radius=2pt];
      \filldraw [black] (b) circle [radius=2pt];
    \end{feynman}
  \end{tikzpicture}
  \caption{Schematic contractions of the six-point effective action}
  \label{fig:contraction-6-point-effective-action}
\end{figure}
The four-point effective action allows two possible contractions. It turns out that the contraction shown on the right in Fig.~\ref{fig:contraction-4-point-effective-action} vanishes even at finite temperature, while the corresponding contraction in Fig.~\ref{fig:contraction-6-point-effective-action} is identically zero due to momentum conservation at the vertex.

Including all Dirac indices, the Rényi entropy can once again be expressed in terms of a set of basketball diagrams, as illustrated in Fig.~\ref{fig:feynman-diagram-renyi-entropy}.
\begin{figure}
  \centering
          \begin{tikzpicture}[baseline=(a.base)]
              \begin{feynman}[every scalar=very thick, every plain=very thick]
                  \vertex (a);
                  \vertex [right=2cm of a] (b);
                  \diagram*{
                  (a) -- [half left, looseness=1.7, anti fermion, edge label=\(\alpha\)] (b),
                  (b) -- [half left, looseness=1.7, charged scalar, edge label=\(\beta\)] (a),
                  (a) -- [half left, looseness=0.8, charged scalar, edge label=\(\beta\)] (b),
                  (b) -- [half left, looseness=0.8, anti charged scalar, edge label'=\(\alpha\)] (a),
                  };
                  \filldraw [black] (a) circle [radius=2pt];
                  \filldraw [black] (b) circle [radius=2pt];
              \end{feynman}
          \end{tikzpicture}
          \quad
          \begin{tikzpicture}[baseline=(a.base)]
              \begin{feynman}[every scalar=very thick, every plain=very thick]
                  \vertex (a);
                  \vertex [right=2cm of a] (b);
                  \diagram*{
                  (a) -- [half left, looseness=1.7, anti fermion, edge label=\(\alpha\)] (b),
                  (b) -- [half left, looseness=1.7, charged scalar, edge label=\(\alpha\)] (a),
                  (a) -- [half left, looseness=0.8, charged scalar, edge label=\(\alpha\)] (b),
                  (b) -- [half left, looseness=0.8, anti charged scalar, edge label'=\(\alpha\)] (a),
                  };
                  \filldraw [black] (a) circle [radius=2pt];
                  \filldraw [black] (b) circle [radius=2pt];
              \end{feynman}
          \end{tikzpicture}
          \\ \quad
          \begin{tikzpicture}[baseline=(a.base)]
              \begin{feynman}[every scalar=very thick, every plain=very thick]
                  \vertex (a);
                  \vertex [right=2cm of a] (b);
                  \diagram*{
                  (a) -- [half left, looseness=1.7, fermion, edge label=\(\beta\)] (b),
                  (b) -- [half left, looseness=1.7, fermion, edge label=\(\beta\)] (a),
                  (a) -- [half left, looseness=0.8, charged scalar, edge label=\(\alpha\)] (b),
                  (b) -- [half left, looseness=0.8, charged scalar, edge label'=\(\alpha\)] (a),
                  };
                  \filldraw [black] (a) circle [radius=2pt];
                  \filldraw [black] (b) circle [radius=2pt];
              \end{feynman}
          \end{tikzpicture}
          \quad
          \begin{tikzpicture}[baseline=(a.base)]
              \begin{feynman}[every scalar=very thick, every plain=very thick]
              \vertex (a);
              \vertex [right=2cm of a] (b);
              \diagram*{
              (a) -- [half left, looseness=1.7, fermion, edge label=\(\alpha\)] (b),
              (b) -- [half left, looseness=1.7, fermion, edge label=\(\alpha\)] (a),
              (a) -- [half left, looseness=0.8, charged scalar, edge label=\(\alpha\)] (b),
              (b) -- [half left, looseness=0.8, charged scalar, edge label'=\(\alpha\)] (a),
              };
              \filldraw [black] (a) circle [radius=2pt];
              \filldraw [black] (b) circle [radius=2pt];
          \end{feynman}
          \end{tikzpicture}
          \\
          \begin{tikzpicture}[baseline=(a.base)]
              \begin{feynman}[every scalar=very thick, every plain=very thick]
                  \vertex (a);
                  \vertex [right=2cm of a] (b);
                  \diagram*{
                  (a) -- [half left, looseness=1.7, fermion, edge label=\(\beta\)] (b),
                  (b) -- [half left, looseness=1.7, fermion, edge label=\(\alpha\)] (a),
                  (a) -- [half left, looseness=0.8, charged scalar, edge label=\(\alpha\)] (b),
                  (b) -- [half left, looseness=0.8, charged scalar, edge label'=\(\beta\)] (a),
                  };
                  \filldraw [black] (a) circle [radius=2pt];
                  \filldraw [black] (b) circle [radius=2pt];
              \end{feynman}
          \end{tikzpicture}
          \quad\begin{tikzpicture}[baseline=(a.base)]
              \begin{feynman}[every scalar=very thick, every plain=very thick]
                  \vertex (a);
                  \vertex [right=2cm of a] (b);
                  \diagram*{
                  (a) -- [half left, looseness=1.7, fermion, edge label=\(\beta\)] (b),
                  (b) -- [half left, looseness=1.7, fermion, edge label=\(\beta\)] (a),
                  (a) -- [half left, looseness=0.8, fermion, edge label=\(\alpha\)] (b),
                  (b) -- [half left, looseness=0.8, charged scalar, edge label'=\(\alpha\)] (a),
                  };
                  \filldraw [black] (a) circle [radius=2pt];
                  \filldraw [black] (b) circle [radius=2pt];
              \end{feynman}
          \end{tikzpicture}
          \quad
          \begin{tikzpicture}[baseline=(a.base)]
              \begin{feynman}[every scalar=very thick, every plain=very thick]
                  \vertex (a);
                  \vertex [right=2cm of a] (b);
                  \diagram*{
                  (a) -- [half left, looseness=1.7, fermion, edge label=\(\alpha\)] (b),
                  (b) -- [half left, looseness=1.7, fermion, edge label=\(\alpha\)] (a),
                  (a) -- [half left, looseness=0.8, fermion, edge label=\(\alpha\)] (b),
                  (b) -- [half left, looseness=0.8, charged scalar, edge label'=\(\alpha\)] (a),
                  };
                  \filldraw [black] (a) circle [radius=2pt];
                  \filldraw [black] (b) circle [radius=2pt];
              \end{feynman}
          \end{tikzpicture}
  \caption{Feynman diagram representation of Rényi entropies}
  \label{fig:feynman-diagram-renyi-entropy}
\end{figure}

So far for the 4 fermion theory the Rényi entropy have been calculated and similar patter as in Refs.~\cite{RN65} 
has been found, for von Neumann entropy it is explained also in Ref.~\cite{RN65} that the replacement of $\frac{n}{n-1}$ by $\text{log}\frac{1}{g^2}$
give the leading order contribution to von Neumann entropies.

\section{conclusions}\label{sec-4}
The methodology established in \onlinecite{RN65} provides a systematic framework for deriving entanglement entropies directly from the Wilsonian effective action. 
This approach offers two major advantages: (1) inherent extensibility to fermionic systems, as demonstrated in our work, and (2) systematic accessibility 
to higher-order corrections, particularly for Rényi entropies. We emphasize that the computational framework involves two distinct perturbative expansions: (1) 
Effective action derivation - Higher-order terms can be systematically incorporated through standard perturbative techniques.(2)
Modified partition function evaluation - Current calculations employ first-order perturbation theory, leaving room for future higher-order extensions.
This dual perturbative structure identifies two critical directions for advancing the formalism.

A key result emerges in the diagrammatic representation of Rényi entropies (Fig.~\ref{fig:feynman-diagram-renyi-entropy}), where "basketball" diagrams encode 
all possible Dirac index contractions and momentum configurations. 

 {More specifically, the two point effective action allows two kinds of contraction of external low-momentum legs.  Their respective contribution to Rényi entropy are denoted by $H_n^{2, 1}(\mu)$ and $H_n^{2, 2}(\mu)$, where} $$\mathcal{T}_{\eta_3\eta_5}(\vk_3,\vk_5)\equiv\text{tr}\;(\Lambda_{\vk_3}^{\eta_3}\gamma^0\Lambda_{\vk_5}^{\eta_5}\gamma^0)$$ and $$\mathcal{T}_{\eta_1\eta_2\eta_3\eta_4}(\vk_1,\vk_2,\vk_3,\vk_4)\equiv\text{tr}(\Lambda_{\vk_1}^{\eta_1}\gamma^0\Lambda_{\vk_2}^{\eta_2}\gamma^0\Lambda_{\vk_3}^{\eta_3}\gamma^0\Lambda_{\vk_4}^{\eta_4}\gamma^0)$$ etc. ($\Lambda_{\vk}^{\eta}$ are projection operators defined in eq. \ref{eq-appedix-profection-operators}):
\begin{equation}
	\begin{aligned}
		&\mathrel{\phantom{=}} H_n^{2,1}(\mu) = \frac{n}{n-1}4g^2\frac{1}{V^2}\sum_{\vk_1\vk_3\vk_4 \vk_5}\delta_{\vk_1+\vk_5,\vk_3+\vk_4}\\
		&\mathrel{\phantom{=}} \times\; \frac{\mathcal{T}_{+-}(\vk_4,\vk_1)\mathcal{T}_{+-}(\vk_3,\vk_5)+\mathcal{T}_{-+}(\vk_4,\vk_1)\mathcal{T}_{-+}(\vk_3,\vk_5)}{(M_1+M_3+M_4+M_5)^2},
	\end{aligned}
\end{equation}

\begin{equation}
	\begin{aligned}
		&\mathrel{\phantom{=}} H_n^{2,2}(\mu) \;=\; -\frac{n}{n-1}4g^2\frac{1}{V^2}\sum_{\vk_1\vk_3\vk_4 \vk_5}\delta_{\vk_1+\vk_5,\vk_3+\vk_4}\\
		&\mathrel{\phantom{=}} \times\; \frac{\mathcal{T}_{-+-+}(\vk_4,\vk_5,\vk_3,\vk_1)+\mathcal{T}_{+-+-}(\vk_4,\vk_5,\vk_3,\vk_1)}{(M_1+M_3+M_4+M_5)^2},\\ 
	\end{aligned}
\end{equation}
 {the Feynman diagrams for these two terms are shown in Fig. \ref{fig-chap05-renyi-entropy-component-2-1} and Fig. \ref{fig-chap05-renyi-entropy-component-2-2} ($1_\alpha$ represent $\vk_{1, \alpha}$ etc.).}
\begin{figure}[!htbp]
	\centering
		\begin{equation*}
		H_n^{2,1}(\mu)
		\quad=\quad\frac{n}{n-1}
		\begin{tikzpicture}[baseline=(a.base)]
			\begin{feynman}[every scalar=very thick, every plain=very thick]
				\vertex (a);
				\vertex [right=2cm of a] (b);
				\diagram*{
				(a) -- [half left, looseness=1.7, anti fermion, edge label={$1_\alpha$}] (b),
				(b) -- [half left, looseness=1.7, charged scalar, edge label=\(5_\beta\)] (a),
				(a) -- [half left, looseness=0.8, charged scalar, edge label=\(\tiny 3_\beta\)] (b),
				(b) -- [half left, looseness=0.8, anti charged scalar, edge label'=\(4_\alpha\)] (a),
				};
				\filldraw [black] (a) circle [radius=2pt];
				\filldraw [black] (b) circle [radius=2pt];
			\end{feynman}
		\end{tikzpicture}
	\end{equation*}
	\caption{Feynman daigram of $ H_n^{2,1}(\mu)$}
	\label{fig-chap05-renyi-entropy-component-2-1}
\end{figure}

\begin{figure}[!htbp]
	\centering
		\begin{equation*}
		H_n^{2,2}(\mu)
		\quad=\quad\frac{n}{n-1}
		\begin{tikzpicture}[baseline=(a.base)]
			\begin{feynman}[every scalar=very thick, every plain=very thick]
				\vertex (a);
				\vertex [right=2cm of a] (b);
				\diagram*{
				(a) -- [half left, looseness=1.7, anti fermion, edge label=\(1_\alpha\)] (b),
				(b) -- [half left, looseness=1.7, charged scalar, edge label=\(5_\alpha\)] (a),
				(a) -- [half left, looseness=0.8, charged scalar, edge label=\(3_\alpha\)] (b),
				(b) -- [half left, looseness=0.8, anti charged scalar, edge label'=\(4_\alpha\)] (a),
				};
				\filldraw [black] (a) circle [radius=2pt];
				\filldraw [black] (b) circle [radius=2pt];
			\end{feynman}
		\end{tikzpicture}
	\end{equation*}
  \caption{Feynman daigram of $ H_n^{2,2}(\mu)$}
	\label{fig-chap05-renyi-entropy-component-2-2}
\end{figure}
 {Note that the only distinction between these two diagrams lies in the structure of the Dirac indices of the propagators; there are no other possible arrangements of these indices. A similar pattern also appears in the contributions from the four-point and six-point effective actions.}

 {Likewise, the four-point effective action yields three distinct contributions to the Rényi entropy, denoted as  $H_{\mu}^{4, 1}(\mu)$, $H_{\mu}^{4, 1}(\mu)$ and $H_{\mu}^{4, 1}(\mu)$ respectively:}
\begin{equation}
	\begin{aligned}
		&\mathrel{\phantom{=}}H_n^{4,1}(\mu) \;=\; \frac{n}{n-1}2g^2\frac{1}{V^2}\sum_{\vk_1\vk_3\vk_5\vk_6}\delta_{\vk_1+\vk_6,\vk_3+\vk_5}\\
		&\mathrel{\phantom{=}} \times\; \frac{\mathcal{T}_{-+}(\vk_1,\vk_3)\mathcal{T}_{+-}(\vk_5,\vk_6)+\mathcal{T}_{+-}(\vk_1,\vk_3)\mathcal{T}_{-+}(\vk_5,\vk_6)}{(M_1+M_3+M_5+M_6)^2}. \\
	\end{aligned}
\end{equation}
The corresponding diagram is shown in Fig.~\ref{fig-chap05-four-fermion-entropy-component-4-1}.
\begin{figure}[!htbp]
	\centering
		\begin{equation*}
		H_n^{4,1}(\mu)
		\quad=\quad\frac{n}{n-1}
		\begin{tikzpicture}[baseline=(a.base)]
			\begin{feynman}[every scalar=very thick, every plain=very thick]
				\vertex (a);
				\vertex [right=2cm of a] (b);
				\diagram*{
				(a) -- [half left, looseness=1.7, fermion, edge label=\(1_\beta\)] (b),
				(b) -- [half left, looseness=1.7, fermion, edge label=\(3_\beta\)] (a),
				(a) -- [half left, looseness=0.8, charged scalar, edge label=\(6_\alpha\)] (b),
				(b) -- [half left, looseness=0.8, charged scalar, edge label'=\(5_\alpha\)] (a),
				};
				\filldraw [black] (a) circle [radius=2pt];
				\filldraw [black] (b) circle [radius=2pt];
			\end{feynman}
		\end{tikzpicture}
	\end{equation*}
  \caption{Feynman daigram of $ H_n^{4,1}(\mu)$}
	\label{fig-chap05-four-fermion-entropy-component-4-1}
\end{figure}

Similarly, we have
\begin{equation}
	\begin{aligned}
		&\mathrel{\phantom{=}}\;H_n^{4,2}(\mu) \;=\; -\frac{n}{n-1}4g^2\frac{1}{V^2}\sum_{\vk_1\vk_3\vk_5\vk_6}\delta_{\vk_1+\vk_6,\vk_3+\vk_5}\\
		&\mathrel{\phantom{=}}\;  \frac{\mathcal{T}_{-+-+}(\vk_3,\vk_6,\vk_5,\vk_1)+\mathcal{T}_{+-+-}(\vk_3,\vk_6,\vk_5,\vk_1)}{(M_1+M_3+M_5+M_6)^2}, \\
	\end{aligned}
\end{equation}
the corresponding diagram is shown shown in Fig. \ref{fig-chap05-four-fermion-entropy-component-4-2}.
\begin{figure}[!htbp]
	\centering
		\begin{equation*}
		H_n^{4,2}(\mu)
		\quad=\quad\frac{n}{n-1}
		\begin{tikzpicture}[baseline=(a.base)]
			\begin{feynman}[every scalar=very thick, every plain=very thick]
				\vertex (a);
				\vertex [right=2cm of a] (b);
				\diagram*{
				(a) -- [half left, looseness=1.7, fermion, edge label=\(1_\alpha\)] (b),
				(b) -- [half left, looseness=1.7, fermion, edge label=\(3_\alpha\)] (a),
				(a) -- [half left, looseness=0.8, charged scalar, edge label=\(6_\alpha\)] (b),
				(b) -- [half left, looseness=0.8, charged scalar, edge label'=\(5_\alpha\)] (a),
				};
				\filldraw [black] (a) circle [radius=2pt];
				\filldraw [black] (b) circle [radius=2pt];
			\end{feynman}
		\end{tikzpicture}
	\end{equation*}
  \caption{Feynman daigram of $ H_n^{4,2}(\mu)$}
	\label{fig-chap05-four-fermion-entropy-component-4-2}
\end{figure}

The third contribution read:
\begin{equation}
	\begin{aligned}
		&\mathrel{\phantom{=}}H_n^{4,3}(\mu) \;=\; \frac{n}{n-1}2g^2\frac{1}{V^2}\sum_{\vk_1\vk_3\vk_5\vk_6}\delta_{\vk_1+\vk_6,\vk_3+\vk_5}\\
		&\mathrel{\phantom{=}} \times\; \frac{\mathcal{T}_{+-}(\vk_5,\vk_1)\mathcal{T}_{+-}(\vk_6,\vk_3)+\mathcal{T}_{-+}(\vk_5,\vk_1)\mathcal{T}_{-+}(\vk_6,\vk_3)}{(M_1+M_3+M_5+M_6)^2},\\
	\end{aligned}
\end{equation}
and the corresponding diagram is shown in Fig. \ref{fig-chap05-four-fermion-entropy-component-4-3}.
\begin{figure}[!htbp]
	\centering
		\begin{equation*}
		H_n^{4,3}(\mu)
		\quad=\quad\frac{n}{n-1}
		\begin{tikzpicture}[baseline=(a.base)]
			\begin{feynman}[every scalar=very thick, every plain=very thick]
				\vertex (a);
				\vertex [right=2cm of a] (b);
				\diagram*{
				(a) -- [half left, looseness=1.7, fermion, edge label=\(1_\beta\)] (b),
				(b) -- [half left, looseness=1.7, fermion, edge label=\(3_\alpha\)] (a),
				(a) -- [half left, looseness=0.8, charged scalar, edge label=\(6_\alpha\)] (b),
				(b) -- [half left, looseness=0.8, charged scalar, edge label'=\(5_\beta\)] (a),
				};
				\filldraw [black] (a) circle [radius=2pt];
				\filldraw [black] (b) circle [radius=2pt];
			\end{feynman}
		\end{tikzpicture}
	\end{equation*}
  \caption{Feynman daigram of $ H_n^{4,3}(\mu)$}
	\label{fig-chap05-four-fermion-entropy-component-4-3}
\end{figure}

For 6-point effective action, we have:
\begin{equation}
	\begin{aligned}
		&\mathrel{\phantom{=}}H_n^6(\mu) \;=\; \frac{n}{n-1}4g^2\frac{1}{V^2}\sum_{\vk_1\vk_2\vk_3 \vk_7}\delta_{\vk_1+\vk_2,\vk_3+\vk_7}\\
    &\mathrel{\phantom{=}} \times \; \frac{1}{(M_1+M_2+M_3+M_7)^2}\\
		&\mathrel{\phantom{=}} \times\; \Big\{\mathcal{T}_{+-}(\vk_3,\vk_1)\mathcal{T}_{+-}(\vk_7,\vk_2)-\mathcal{T}_{+-+-}(\vk_3,\vk_2,\vk_7,\vk_1) \\
		&\mathrel{\phantom{+}}\; +\; \mathcal{T}_{-+}(\vk_3,\vk_1)\mathcal{T}_{-+}(\vk_7,\vk_2)-\mathcal{T}_{-+-+}(\vk_3,\vk_2,\vk_7,\vk_1) \Big\}\\
	\end{aligned}
\end{equation}
and its diagram is shown in Fig. \ref{fig-chap05-four-feynman-6}.
\begin{figure}[!htbp]
	\centering
		\begin{equation*}
		H_n^{6}(\mu)
		=\frac{n}{n-1}
		\begin{tikzpicture}[baseline=(a.base)]
			\begin{feynman}[every scalar=very thick, every plain=very thick]
				\vertex (a);
				\vertex [right=2cm of a] (b);
				\diagram*{
				(a) -- [half left, looseness=1.7, fermion, edge label=\(\beta\)] (b),
				(b) -- [half left, looseness=1.7, fermion, edge label=\(\beta\)] (a),
				(a) -- [half left, looseness=0.8, fermion, edge label=\(\alpha\)] (b),
				(b) -- [half left, looseness=0.8, charged scalar, edge label'=\(\alpha\)] (a),
				};
				\filldraw [black] (a) circle [radius=2pt];
				\filldraw [black] (b) circle [radius=2pt];
			\end{feynman}
		\end{tikzpicture}
    +\frac{n}{n-1}
    \begin{tikzpicture}[baseline=(a.base)]
			\begin{feynman}[every scalar=very thick, every plain=very thick]
				\vertex (a);
				\vertex [right=2cm of a] (b);
				\diagram*{
				(a) -- [half left, looseness=1.7, fermion, edge label=\(\alpha\)] (b),
				(b) -- [half left, looseness=1.7, fermion, edge label=\(\alpha\)] (a),
				(a) -- [half left, looseness=0.8, fermion, edge label=\(\alpha\)] (b),
				(b) -- [half left, looseness=0.8, charged scalar, edge label'=\(\alpha\)] (a),
				};
				\filldraw [black] (a) circle [radius=2pt];
				\filldraw [black] (b) circle [radius=2pt];
			\end{feynman}
		\end{tikzpicture}
	\end{equation*}
  \caption{Feynman daigram of $ H_n^{6}(\mu)$}
	\label{fig-chap05-four-feynman-6}
\end{figure}

 {Note that although we have specifically partitioned the momentum space into two subsets—above and below an arbitrary scale $\mu$—our method and the resulting expressions are valid for any partition of the momentum space. Moreover, compared to the application of this method to the bosonic case in Ref.~\cite{RN65}, we refine the calculation by taking the zero-temperature limit at the final step, as detailed in Appendix~\ref{appendix}, rather than at an intermediate step as done in Ref.~\cite{RN65}. This procedure renders our final results more reliable and facilitates their generalization to higher-order calculations.}

\begin{figure}
    \centering
    \includegraphics[width=.45\textwidth]{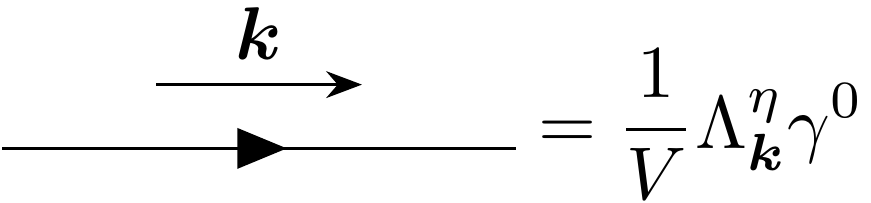}
    \caption{Fermion propagator}
    \label{fig:propagator-fermion} 
\end{figure}

\begin{figure}
    \centering
    \includegraphics[width=.45\textwidth]{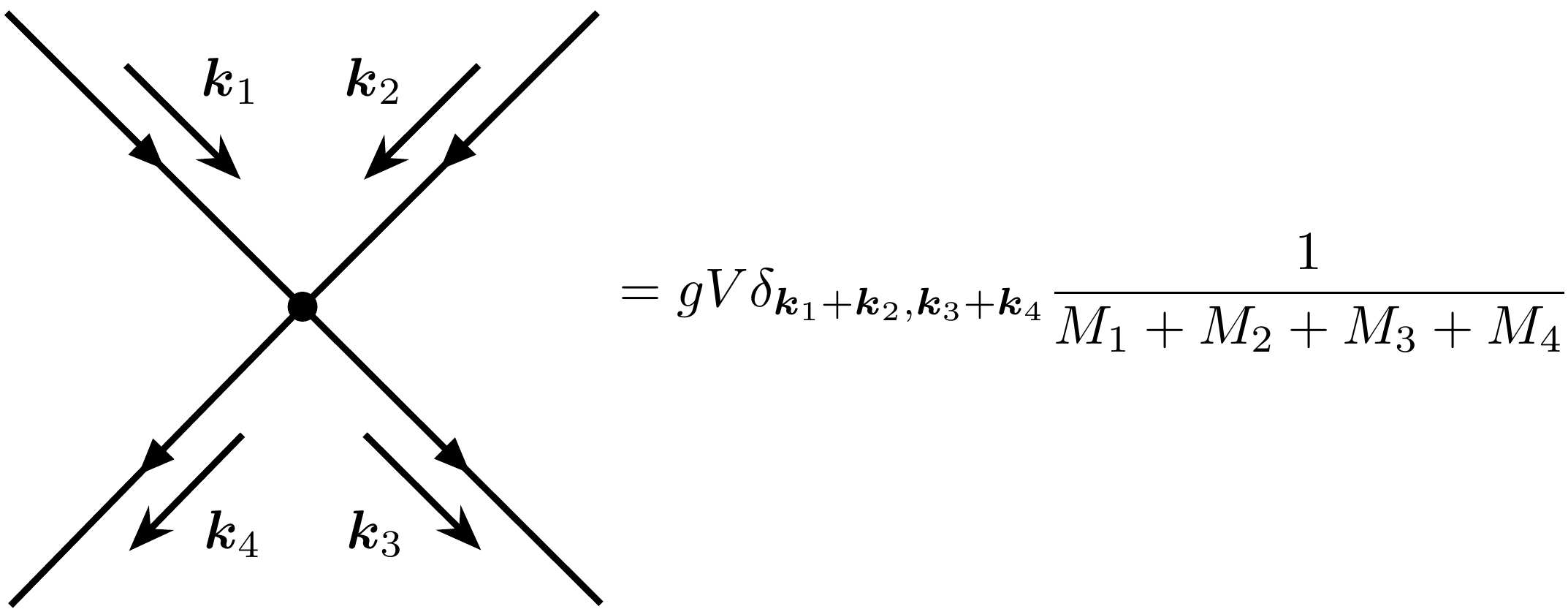}
    \caption{4-fermion vertex}
    \label{fig:vertex-fermion}
\end{figure}

Finally, to interpret these structures, we propose modified Feynman rules (Figs.~\ref{fig:propagator-fermion}-\ref{fig:vertex-fermion}) 
incorporating a novel sign alternation rule for adjacent projection operators – a distinctive feature rooted in the Grassmannian nature of fermionic path integrals (see Appendix~\ref{appendix}).
 While this sign alternation mechanism lacks a complete theoretical interpretation. This empirical consistency across both bosonic and fermionic systems suggests a universal characteristic of momentum-space entanglement.
The phenomenological success of these diagrammatic rules raises fundamental questions: Are their structural similarities between scalar and fermionic theories merely coincidental, 
or do they reflect deeper principles governing momentum-mode entanglement? Resolving this ambiguity requires investigating of higher-order corrections to test wether the observed patterns persist or not.

These open questions underscore the need for a unified theoretical framework capable of explaining the microscopic origin of entanglement signatures across diverse quantum field theories.



\begin{acknowledgments}
We wish to acknowledge Maoling Zhou for useful disscussions.
\end{acknowledgments}

\appendix
\onecolumngrid
\section{MOMENTUM-SPACE ENTROPY IN $(\bar{\psi}\psi)^2$ THEORY}\label{appendix}
We first give explicitly the Feynman rules read from the Euclidean actions as follows,
\begin{enumerate}
  \item enumerate all topologically equivelent diagram;
  \item dashed lines represent high momentum modes and solid lines represent low-momentum modes;
  \item fermion propagators are given by $\frac{1}{\beta}\mathcal{S}_0(K)$;
  \item the 4 fermion vertex is given by $-g\frac{\beta}{V}$;
  \item momentum conservation $\beta V\delta_{K_i,K_f}$ at each vertex where $K_i$ is the sum of incoming momentum and $K_f$ is the sum of outcomming momentum;
  \item external legs are $\psi_{0,\vk}$ and $\bar{\psi}_{0,\vk}$;
  \item momentum integral for high-momentum modes in the range $[\mu, \Lambda]$ and momentum integral for low-momentum modes in the range $[0, \mu]$;
  \item extra minus sign due to fermion loop.
\end{enumerate}

Fig.~\ref{fig:order1fermion} is a local contribution which does not contribute so we will not consider this term. We define the two kinds of 2-point effective action as $S_{\mu}^{\beta,(2,1)}$
and $S_{\mu}^{\beta,(2,2)}$, define the three kinds of 4-point effective action as $S_{\mu}^{\beta,(4,1)},S_{\mu}^{\beta,(4,2)},S_{\mu}^{\beta,(4,3)}$ and the 6-point effective action as $S_{\mu}^{\beta,(6)}$ , then from the Feynman rules above we can write down the expressions explicitly,
	\begin{align}
		\mathrel{\phantom{=}}\; S_{\mu}^{\beta,(2,1)} 
		&=\; -\frac{1}{2}\times2\times4\times\left(g\frac{\beta}{V}\right)^2\sum_{K_1K_2K_3K_4K_5}(-1)\text{tr}\left(\frac{\mathcal{S}_0(K_3)}{\beta}\frac{\mathcal{S}_0(K_5)}{\beta}\right)\bar{\psi}_{0,K_1}\frac{\mathcal{S}_0(K_4)}{\beta}\psi_{0,K_2} \nonumber\\
		&\mathrel{\phantom{=}}\; \times\; \delta_{K_1+K_5-K_3-K_4, 0}\cdot\delta_{K_3+K_4-K_2-K_5, 0}\\
		\mathrel{\phantom{=}}\; S_{\mu}^{\beta,(2,2)} 
		&=\; -\frac{1}{2}\times2\times4\times\left(g\frac{\beta}{V}\right)^2\sum_{K_1K_2K_3K_4K_5}\bar{\psi}_{0,K_1}\frac{\mathcal{S}_0(K_4)}{\beta}\frac{\mathcal{S}_0(K_5)}{\beta}\frac{\mathcal{S}_0(K_3)}{\beta}\psi_{0,K_2}\nonumber\\
		&\mathrel{\phantom{=}}\; \times\; \delta_{K_1+K_5-K_3-K_4, 0}\cdot\delta_{K_3+K_4-K_2-K_5, 0}\\
		\mathrel{\phantom{=}}\; S_{\mu}^{\beta,(4,1)} 
		&=\;  -\frac{2\times2}{2}\times\left(g\frac{\beta}{V}\right)^2\sum_{K_1K_2K_3K_4K_5K_6}\bar{\psi}_{0,K_1}\psi_{0,K_2}(-1)\text{tr}\left(\frac{\mathcal{S}_0(K_5)}{\beta}\frac{\mathcal{S}_0(K_6)}{\beta}\right)\bar{\psi}_{0,K_3}\psi_{0,K_4}\nonumber\\
		&\mathrel{\phantom{=}}\; \times\; \delta_{K_1+K_6-K_2-K_5, 0}\cdot\delta_{K_3+K_5-K_6-K_4, 0}\\
		\mathrel{\phantom{=}}\; S_{\mu}^{\beta,(4,2)} 
		&=\;  -\frac{2\times4}{2}\times\left(g\frac{\beta}{V}\right)^2\sum_{K_1K_2K_3K_4K_5K_6}\bar{\psi}_{0,K_1}\psi_{0,K_2}\bar{\psi}_{0,K_3}\left(\frac{\mathcal{S}_0(K_6)}{\beta}\frac{\mathcal{S}_0(K_5)}{\beta}\right)\psi_{0,K_4}\nonumber\\
		&\mathrel{\phantom{=}}\; \times\; \delta_{K_1+K_6-K_2-K_5, 0}\cdot\delta_{K_3+K_5-K_6-K_4, 0}\\
		\mathrel{\phantom{=}}\; S_{\mu}^{\beta,(4,3)} 
		&=\;  -\frac{2\times2}{2}\times\left(g\frac{\beta}{V}\right)^2\sum_{K_1K_2K_3K_4K_5K_6}\bar{\psi}_{0,K_1}\frac{\mathcal{S}_0(K_5)}{\beta}\psi_{0,K_4}\bar{\psi}_{0,K_3}\frac{\mathcal{S}_0(K_6)}{\beta}\psi_{0,K_2}\nonumber\\
		&\mathrel{\phantom{=}}\; \times\; \delta_{K_1+K_6-K_2-K_5, 0}\cdot\delta_{K_3+K_5-K_6-K_4, 0}\\
		\mathrel{\phantom{=}}\; S_{\mu}^{\beta,(6)} 
		&=\;  -\frac12\times2\times4\times\left(g\frac{\beta}{V}\right)^2\sum_{K_1K_2K_3K_4K_5K_6K_7}\bar{\psi}_{0,K_1}\psi_{0,K_3}\bar{\psi}_{0,K_4}\psi_{0,K_6}\bar{\psi}_{0,K_2}\frac{\mathcal{S}_0(K_7)}{\beta}\psi_{0,K_5}\nonumber\\
		&\mathrel{\phantom{=}}\; \times\; \delta_{K_1+K_2-K_3-K_7, 0}\cdot\delta_{K_7+K_4-K_5-K_6, 0}\\
	\end{align}

The above expressions for effective actions are in momentum space, in order to find the "glued" effective action, we'd better transform the above expressions in Euclidean time formalism in which 
the mixed representation of propagators can be used to finish Matsubara summations.

For that purposes we employ the following transformations of low-momentum modes,
\begin{equation}\label{eq-chap05-mixed-field}
	\begin{aligned}
		\psi_{0,K} \;&=\; T\int_{0}^{\beta}\d\tau \e^{-\i\omega_j\tau}\psi_{0,\vk}(\tau), \quad\quad \psi_{0,\vk}(\tau)=\sum_{j}\e^{\i\omega_j\tau}\psi_{0,K}. \\
		\bar{\psi}_{0,K} \;&=\; T\int_{0}^{\beta}\d\tau \e^{\i\omega_j\tau}\bar{\psi}_{0,\vk}(\tau), \quad\quad \bar{\psi}_{0,\vk}(\tau)=\sum_{j}\e^{-\i\omega_j\tau}\bar{\psi}_{0,K}. \\
	\end{aligned}
\end{equation}
Besides, let's review the mixed representation of fermions propagators, define projection operators as follows,
\begin{equation}\label{eq-appedix-profection-operators}
	\begin{aligned}
		\Lambda_{\vk}^{\pm} \;\equiv\; \frac{M\pm\gamma^0(\bm{\gamma}\cdot\vk+m)}{2M}
	\end{aligned}
\end{equation}
then the thermal fermion propagator can be writen as
\begin{equation}\label{eq-appendix-fermion-propagator-decomposition}
	\begin{aligned}
		\S_0(K) \;&=\; \frac{k_0\gamma^0-\bm{\gamma}\cdot\vk+m}{(M-k_0)(M+k_0)} \\
		&=\; -\sum_{\eta=\pm}\frac{1}{k_0-\eta M}\Lambda_{\vk}^{\eta}\gamma^0 \\
		&=\; \sum_{\eta=\pm}\S^{\eta}_0(K)\Lambda_{\vk}^{\eta}\gamma^0
	\end{aligned}
\end{equation}
Then we can define the mixed representation of fermion propagator as
\begin{equation}\label{eq-appendix-fermion-mixed-propagator}
	\begin{aligned}
		\bar{\S}_0^{\eta}(\tau, M) \;&=\; T\sum_j \e^{\i\omega_j\tau}\S_0^{\eta}(K), \\
		\S_0^{\eta}(K) \;&=\; \int_{0}^{\beta}\d\tau \;\e^{-\i\omega_j\tau}\bar{\S}_0^{\eta}(\tau, M) \\
	\end{aligned}
\end{equation}
where 
\begin{equation}\label{eq-appendix-fermion-mixed-propagator-explicit}
	\bar{\S}_0^{\eta}(\tau, M) \;=\; \e^{-\eta M\tau}[1-n_F(\eta M)]
\end{equation}

Substitute the mixed representation into effective action $S_{\mu}^{\beta,(2,1)}$, we get
\begin{equation}
	\begin{aligned}
		\mathrel{\phantom{=}}\; S_{\mu}^{\beta,(2,1)} 
		&=\; 4g^2\frac{1}{\beta V^2}\sum_{K_1\to K_5}\delta_{K_1+K_5-K_3-K_4, 0}\cdot\delta_{K_3+K_4-K_2-K_5, 0} \sum_{\eta_3\eta_4\eta_5}\S_0^{\eta_3}(K_3)\S_0^{\eta_4}(K_4)\S_0^{\eta_5}(K_5)\\
		&\mathrel{\phantom{=}}\; \times\; \text{tr}\;(\Lambda_{\vk_3}^{\eta_3}\gamma^0\Lambda_{\vk_5}^{\eta_5}\gamma^0) \;\bar{\psi}_{0,K_1}\Lambda_{\vk_4}^{\eta_4}\gamma^0\psi_{0,K_2} 
	\end{aligned}
\end{equation}
define $\mathcal{T}_{\eta_3\eta_5}(\vk_3,\vk_5)\equiv\text{tr}\;(\Lambda_{\vk_3}^{\eta_3}\gamma^0\Lambda_{\vk_5}^{\eta_5}\gamma^0)$, then substitute the transformation Eq.~\ref{eq-chap05-mixed-field} in, we have
\begin{equation*}
	\begin{aligned}
		\mathrel{\phantom{=}}\; S_{\mu}^{\beta,(2,1)} 
		&=\; 4g^2\frac{1}{\beta^3 V^2}\sum_{K_1\to K_5}\delta_{K_1+K_5-K_3-K_4,0}\cdot\delta_{K_3+K_4-K_2-K_5,0}\sum_{\eta_3\eta_4\eta_5}\int_{0}^{\beta}\d\tau_1\cdots\d\tau_5 \\
		&\mathrel{\phantom{=}}\; \times\; \bar{\S}_0^{\eta_3}(\tau_3,  M_3)\bar{\S}_0^{\eta_4}(\tau_4, M_4)\bar{\S}_0^{\eta_5}(\tau_5, M_5)\mathcal{T}_{\eta_3\eta_5}(\vk_3,\vk_5)\bar{\psi}_{0,\vk_1}(\tau_1)\Lambda_{\vk_4}^{\eta_4}\gamma^0\psi_{0,\vk_2}(\tau_2) \\
		&\mathrel{\phantom{=}}\; \times\; \e^{\i\omega_{j_1}\tau_1-\i\omega_{j_2}\tau_2-\i\omega_{j_3}\tau_3-\i\omega_{j_4}\tau_4-\i\omega_{j_5}\tau_5} \\
	\end{aligned}
\end{equation*}
Now the Matsubara frenquencies summations can be easily performed, and we arrive at a formula that is suitiable for replica tricks as follows,
\begin{equation*}
	\begin{aligned}
		S_{\mu}^{\beta,(2,1)}  \;&=\; 4g^2\frac{1}{V^2}\sum_{\vk_1\to \vk_5}\delta_{\vk_1+\vk_5,\vk_3+\vk_4}\cdot\delta_{\vk_1,\vk_2}\sum_{\eta_3\eta_4\eta_5}(I_1^{(2,1)}+I_2^{(2,1)})\\
	\end{aligned}
\end{equation*}
where
\begin{equation*}
	\begin{aligned}
		I_1^{(2,1)} &\equiv -\int_{0}^{\beta}\d\tau_1\int_{0}^{\tau_1}\d\tau_2\bar{\S}_0^{\eta_3}(\tau_1-\tau_2, M_3)\bar{\S}_0^{\eta_4}(\tau_1-\tau_2, M_4)\bar{\S}_0^{-\eta_5}(\tau_1-\tau_2, M_5)\mathcal{T}_{\eta_3\eta_5}(\vk_3,\vk_5) \\
        &\mathrel{\phantom{=}}\; \times\; \bar{\psi}_{0,\vk_1}(\tau_1)\cdot\Lambda_{\vk_4}^{\eta_4}\gamma^0\cdot\psi_{0,\vk_2}(\tau_2) \\ 
		I_2^{(2,1)} &\equiv \int_{0}^{\beta}\d\tau_1\int_{\tau_1}^{\beta}\d\tau_2\bar{\S}_0^{-\eta_3}(\tau_2-\tau_1, M_3)\bar{\S}_0^{-\eta_4}(\tau_2-\tau_1, M_4)\bar{\S}_0^{\eta_5}(\tau_2-\tau_1, M_5)\mathcal{T}_{\eta_3\eta_5}(\vk_3,\vk_5)\\
        &\mathrel{\phantom{=}}\; \times\; \bar{\psi}_{0,\vk_1}(\tau_1)\cdot\Lambda_{\vk_4}^{\eta_4}\gamma^0\cdot\psi_{0,\vk_2}(\tau_2) \\ 
	\end{aligned}
\end{equation*}
In the same vein we have the expressions for other effective actions as follows,
\begin{equation}
	\begin{aligned}
		S_{\mu}^{\beta,(2,2)}  \;&=\; -4g^2\frac{1}{V^2}\sum_{\vk_1\to \vk_5}\delta_{\vk_1+\vk_5,\vk_3+\vk_4}\cdot\delta_{\vk_1,\vk_2}\sum_{\eta_3\eta_4\eta_5}\{I_1^{(2,2)}+I_2^{(2,2)}\}\\
	\end{aligned}
\end{equation}
where
\begin{equation}
	\begin{aligned}
		I_1^{(2,2)} &\equiv -\int_{0}^{\beta}\d\tau_1\int_{0}^{\tau_1}\d\tau_2\bar{\psi}_{0,\vk_1}(\tau_1)\Lambda_{\vk_4}^{\eta_4}\gamma^0\Lambda_{\vk_5}^{\eta_5}\gamma^0\Lambda_{\vk_3}^{\eta_3}\gamma^0\psi_{0,\vk_2}(\tau_2) \\
        &\mathrel{\phantom{=}}\; \times\; \bar{\S}_0^{\eta_3}(\tau_1-\tau_2, M_3)\bar{\S}_0^{\eta_4}(\tau_1-\tau_2, M_4)\bar{\S}_0^{-\eta_5}(\tau_1-\tau_2, M_5)\\
		I_2^{(2,2)} &\equiv \int_{0}^{\beta}\d\tau_1\int_{\tau_1}^{\beta}\d\tau_2\bar{\psi}_{0,\vk_1}(\tau_1)\Lambda_{\vk_4}^{\eta_4}\gamma^0\Lambda_{\vk_5}^{\eta_5}\gamma^0\Lambda_{\vk_3}^{\eta_3}\gamma^0\psi_{0,\vk_2}(\tau_2) \\
        &\mathrel{\phantom{=}}\; \times\; \bar{\S}_0^{-\eta_3}(\tau_2-\tau_1, M_3)\bar{\S}_0^{-\eta_4}(\tau_2-\tau_1, M_4)\bar{\S}_0^{\eta_5}(\tau_2-\tau_1, M_5) \\
	\end{aligned}
\end{equation}
The expression for $S_{\mu}^{\beta,(4,1)}$ read
\begin{equation}
	\begin{aligned}
		S_{\mu}^{\beta,(4,1)} \;&=\; 2\frac{g^2}{V^2}\sum_{\vk_1\to \vk_6}\delta_{\vk_1+\vk_6,\vk_2+\vk_5}\delta_{\vk_3+\vk_5,\vk_4+\vk_6}\sum_{\eta_5\eta_6}\{I_1^{(4,1)}+I_2^{(4,1)}\}
	\end{aligned}
\end{equation}
where
\begin{equation}
	\begin{aligned}
		I_1^{(4,1)} &\equiv -\int_0^{\beta}\d\tau_1\int_0^{\tau_1}\d\tau_4 \bar{\psi}_{0,\vk_1}(\tau_1)\psi_{0,\vk_2}(\tau_1)\bar{\psi}_{0,\vk_3}(\tau_4)\psi_{0,\vk_4}(\tau_4)\mathcal{T}_{\eta_5\eta_6}(\vk_5,\vk_6)\\
		&\mathrel{\phantom{=}}\; \times\; \bar{\S}_0^{\eta_5}(\tau_1-\tau_4, M_5)\bar{\S}_0^{-\eta_6}(\tau_1-\tau_4, M_6)\\ 
		I_2^{(4,1)} &\equiv -\int_0^{\beta}\d\tau_1\int_{\tau_1}^{\beta}\d\tau_4 \bar{\psi}_{0,\vk_1}(\tau_1)\psi_{0,\vk_2}(\tau_1)\bar{\psi}_{0,\vk_3}(\tau_4)\psi_{0,\vk_4}(\tau_4)\mathcal{T}_{\eta_5\eta_6}(\vk_5,\vk_6)\\
		&\mathrel{\phantom{=}}\; \times\; \bar{\S}_0^{-\eta_5}(\tau_4-\tau_1, M_5)\bar{\S}_0^{\eta_6}(\tau_4-\tau_1, M_6)
	\end{aligned}
\end{equation}
The expression for $S_{\mu}^{\beta,(4,2)}$ read
\begin{equation}
	\begin{aligned}
		S_{\mu}^{\beta,(4,2)} \;&=\;4g^2\frac{1}{V^2}\sum_{\vk_1\to\vk_6}\delta_{\vk_1+\vk_6,\vk_2+\vk_5}\delta_{\vk_3+\vk_5,\vk_4+\vk_6}\sum_{\eta_5\eta_6}\{I_1^{(4,2)}+I_2^{(4,2)}\}
	\end{aligned}
\end{equation}
where 
\begin{equation}
	\begin{aligned}
		I_1^{(4,2)} &= \int_{0}^{\beta}\d\tau_1\int_{0}^{\tau_1}\d\tau_4\bar{\psi}_{0,\vk_1}(\tau_1)\psi_{0,\vk_2}(\tau_1)\bar{\psi}_{0,\vk_3}(\tau_4)\Lambda_{\vk_6}^{\eta_6}\gamma^0\Lambda_{\vk_5}^{\eta_5}\gamma^0\psi_{0,\vk_4}(\tau_4)\\
		&\mathrel{\phantom{=}}\; \times\; \bar{\S}_0^{-\eta_6}(\tau_1-\tau_4, M_6)\bar{\S}_0^{\eta_5}(\tau_1-\tau_4, M_5)\\
		I_2^{(4,2)} &= \int_{0}^{\beta}\d\tau_1\int_{\tau_1}^{\beta}\d\tau_4\bar{\psi}_{0,\vk_1}(\tau_1)\psi_{0,\vk_2}(\tau_1)\bar{\psi}_{0,\vk_3}(\tau_4)\Lambda_{\vk_6}^{\eta_6}\gamma^0\Lambda_{\vk_5}^{\eta_5}\gamma^0\psi_{0,\vk_4}(\tau_4)\\
		&\mathrel{\phantom{=}}\; \times\; \bar{\S}_0^{\eta_6}(\tau_4-\tau_1, M_6)\bar{\S}_0^{-\eta_5}(\tau_4-\tau_1, M_5)\\
	\end{aligned}
\end{equation}
The expression for $S_{\mu}^{\beta,(4,3)}$ read
\begin{equation}
	\begin{aligned}
		S_{\mu}^{\beta,(4,3)} \;&=\; 2g^2\frac{1}{V^2}\sum_{\vk_1\to\vk_6}\delta_{\vk_1+\vk_6,\vk_2+\vk_5}\delta_{\vk_3+\vk_5,\vk_4+\vk_6}\sum_{\eta_5\eta_6}\{I_1^{(4,3)}+I_2^{(4,3)}\}
	\end{aligned}
\end{equation}
where
\begin{equation}
	\begin{aligned}
		I_1^{(4,3)} &\equiv \int_{0}^{\beta}\d\tau_1\int_{0}^{\tau_1}\d\tau_4\bar{\psi}_{0,\vk_1}(\tau_1)\Lambda_{\vk_5}^{\eta_5}\gamma^0\psi_{0,\vk_4}(\tau_4)\bar{\psi}_{0,\vk_3}(\tau_4)\Lambda_{\vk_6}^{\eta_6}\gamma^0\psi_{0,\vk_2}(\tau_1)\\
		&\mathrel{\phantom{=}} \times\; \bar{\S}_0^{-\eta_6}(\tau_1-\tau_4, M_6)\bar{\S}_0^{\eta_5}(\tau_1-\tau_4, M_5)\\
		I_2^{(4,3)} &\equiv \int_{0}^{\beta}\d\tau_1\int_{\tau_1}^{\beta}\d\tau_4\bar{\psi}_{0,\vk_1}(\tau_1)\Lambda_{\vk_5}^{\eta_5}\gamma^0\psi_{0,\vk_4}(\tau_4)\bar{\psi}_{0,\vk_3}(\tau_4)\Lambda_{\vk_6}^{\eta_6}\gamma^0\psi_{0,\vk_2}(\tau_1)\\
		&\mathrel{\phantom{=}} \times\; \bar{\S}_0^{\eta_6}(\tau_4-\tau_1, M_6)\bar{\S}_0^{-\eta_5}(\tau_4-\tau_1, M_5)
	\end{aligned}
\end{equation}
The expression for $S_{\mu}^{\beta,(6)}$ read
\begin{equation}
	\begin{aligned}
		S_{\mu}^{\beta,(6)} \;&=\; -4g^2\frac{1}{V^2}\sum_{\vk_1\to \vk_7}\delta_{\vk_1+\vk_2,\vk_3+\vk_7}\cdot\delta_{\vk_7+\vk_4,\vk_5+\vk_6}\sum_{\eta_7}\{I_1^{(6)}+I_2^{(6)}\}\\
	\end{aligned}
\end{equation}
where
\begin{equation}
	\begin{aligned}
		I_1^{(6)} &\equiv \int_{0}^{\beta}\d\tau_2\int_{0}^{\tau_2}\d\tau_5 \bar{\psi}_{0,\vk_1}(\tau_2)\psi_{0,\vk_3}(\tau_2)\bar{\psi}_{0,\vk_4}(\tau_5)\psi_{0,\vk_6}(\tau_5)\bar{\psi}_{0,\vk_2}(\tau_2)\Lambda_{\vk_7}^{\eta_7}\gamma^0\psi_{0,\vk_5}(\tau_5)\\
		&\mathrel{\phantom{=}} \times\; \bar{\S}_0^{\eta_7}(\tau_2-\tau_5,\vk_7)\\
		I_2^{(6)} &\equiv -\int_{0}^{\beta}\d\tau_2\int_{\tau_2}^{\beta}\d\tau_5 \bar{\psi}_{0,\vk_1}(\tau_2)\psi_{0,\vk_3}(\tau_2)\bar{\psi}_{0,\vk_4}(\tau_5)\psi_{0,\vk_6}(\tau_5)\bar{\psi}_{0,\vk_2}(\tau_2)\Lambda_{\vk_7}^{\eta_7}\gamma^0\psi_{0,\vk_5}(\tau_5)\\
		&\mathrel{\phantom{=}} \times\; \bar{\S}_0^{-\eta_7}(\tau_5-\tau_2,\vk_7)
	\end{aligned}
\end{equation}
Then we can proceed to calculate the "glued" effective action, the essential part is the following sum of integrations,
\begin{equation}\label{eq-appendix-fermion-replica-integral}
	\begin{aligned}
		A_1(\omega_j,\omega_{j'},M) \;&=\; \frac{1}{n\beta}\sum_{u=0}^{n-1}\int_{u\beta}^{(u+1)\beta}\d\tau\int_{u\beta}^{\tau}\d\tau'\e^{\i\omega_j\tau+\i\omega_{j'}\tau'}\e^{-M(\tau-\tau')}\\
		&=\; \delta_{j+j'}^0\frac{1}{M+\i\omega_{j'}} + \frac{1}{n\beta}\cdot n\sum_\nu\delta_{j+j'}^{n\nu}\frac{\e^{(\i\omega_j-M)\beta}-1}{(M+\i\omega_{j'})(M-\i\omega_j)}\\
		&\equiv\; B_1(\omega_j,\omega_{j'},M)+C_1(\omega_j,\omega_{j'},M)\\
		A_2(\omega_j,\omega_{j'},M) \;&=\; \frac{1}{n\beta}\sum_{u=0}^{n-1}\int_{u\beta}^{(u+1)\beta}\d\tau\int_{\tau}^{(u+1)\beta}\d\tau'\e^{\i\omega_j\tau+\i\omega_{j'}\tau'}\e^{-M(\tau'-\tau)}\\
		&=\; \delta_{j+j'}^0\frac{1}{M-\i\omega_{j'}} + \frac{1}{n\beta}\cdot n\sum_\nu\delta_{j+j'}^{n\nu}\frac{\e^{(\i\omega_{j'}-M)\beta}-1}{(M+\i\omega_{j})(M-\i\omega_{j'})}\\
		&\equiv\; B_2(\omega_j,\omega_{j'},M)+C_2(\omega_j,\omega_{j'},M)\\
	\end{aligned}
\end{equation}
In the above formula we have define the local part as B and nonlocal part as C and we will see that the part B in "glued" effective action coincides with the oringinal effective action at temperture $n\beta$.

Recall that after replica, the low-momentum modes are anti-periodic in the Euclidean time range $[0, n\beta]$, so we can transform the low-momentum modes as,
\begin{equation}
	\begin{aligned}
		\psi_{0,\vk}(\tau)=\sum_{j}\e^{\i\omega_j\tau}\psi_{0,K}, \qquad \bar{\psi}_{0,\vk}(\tau)=\sum_{j}\e^{-\i\omega_j\tau}\bar{\psi}_{0,K}. \\
	\end{aligned}
\end{equation}
now we have $\omega_j=\frac{(2j+1)\pi}{n\beta}$, using these tools, we arrive at the form of the "glued" effective action.

The expression for $S_{\mu,\r}^{\beta,(2,1)}$ read,
\begin{equation}\label{eq-chap05-effective-action-replica-2-1}
	\begin{aligned}
		S_{\mu,\r}^{\beta,(2,1)} \equiv 4g^2\frac{1}{V^2}\sum_{\vk_1\to \vk_5}\delta_{\vk_1+\vk_5,\vk_3+\vk_4}\cdot\delta_{\vk_1,\vk_2}\sum_{\eta_3\eta_4\eta_5}(I_{1,\r}^{(2,1)}+I_{2,\r}^{(2,1)})
	\end{aligned}
\end{equation}
where
\begin{equation}
	\begin{aligned}
		I_{1, \r}^{(2,1)} &= -\sum_{j_1,j_2}\bar{\psi}_{0,K_1}\Lambda_{\vk_4}^{\eta_4}\gamma^0\psi_{0,K_2}\mathcal{T}_{\eta_3\eta_5}(\vk_3,\vk_5)[1-n_F(\eta_3M_3)][1-n_F(\eta_4M_4)]n_F(\eta_5M_5)\\
		&\mathrel{\phantom{=}}\;  \times\; (n\beta) A_1(-\omega_{j_1},\omega_{j_2},\eta_3M_3+\eta_4M_4-\eta_5M_5)\\
		I_{2, \r}^{(2,1)} &= \sum_{j_1,j_2}\bar{\psi}_{0,K_1}\Lambda_{\vk_4}^{\eta_4}\gamma^0\psi_{0,K_2}\mathcal{T}_{\eta_3\eta_5}(\vk_3,\vk_5)n_F(\eta_3M_3)n_F(\eta_4M_4)[1-n_F(\eta_5M_5)]\\
		&\mathrel{\phantom{=}}\;  \times\; (n\beta)A_2(-\omega_{j_1},\omega_{j_2},-\eta_3M_3-\eta_4M_4+\eta_5M_5)
	\end{aligned}
\end{equation}
The expression for $S_{\mu,\r}^{\beta,(2,2)}$ read,
\begin{equation}\label{eq-chap05-effective-action-replica-2-2}
	S_{\mu, \r}^{\beta,(2,2)} \equiv-4g^2\frac{1}{V^2}\sum_{\vk_1\to \vk_5}\delta_{\vk_1+\vk_5,\vk_3+\vk_4}\cdot\delta_{\vk_1,\vk_2}\sum_{\eta_3\eta_4\eta_5}(I_{1,\r}^{(2,2)}+I_{2,\r}^{(2,2)})\\
\end{equation}
where
\begin{equation}
	\begin{aligned}
		I_{1,\r}^{(2,2)} &\equiv -\sum_{j_1j_2}\bar{\psi}_{0,K_1}\Lambda_{\vk_4}^{\eta_4}\gamma^0\Lambda_{\vk_5}^{\eta_5}\gamma^0\Lambda_{\vk_3}^{\eta_3}\gamma^0\psi_{0,K_2}[1-n_F(\eta_3M_3)][1-n_F(\eta_4M_4)]n_F(\eta_5M_5) \\
		&\mathrel{\phantom{=}}\; \times\; (n\beta)A_1(-\omega_{j_1},\omega_{j_2},\eta_3M_3+\eta_4M_4-\eta_5M_5)\\
		I_{2,\r}^{(2,2)} &\equiv \sum_{j_1j_2}\bar{\psi}_{0,K_1}\Lambda_{\vk_4}^{\eta_4}\gamma^0\Lambda_{\vk_5}^{\eta_5}\gamma^0\Lambda_{\vk_3}^{\eta_3}\gamma^0\psi_{0,K_2}n_F(\eta_3M_3)n_F(\eta_4M_4)[1-n_F(\eta_5M_5)] \\
		&\mathrel{\phantom{=}}\; \times\;  (n\beta)A_2(-\omega_{j_1},\omega_{j_2},-\eta_3M_3-\eta_4M_4+\eta_5M_5)
	\end{aligned}
\end{equation}
The expression for $S_{\mu,\r}^{\beta,(4,1)}$ read,
\begin{equation}\label{eq-chap05-effective-action-replica-4-1}
	\begin{aligned}
		S_{\mu,\r}^{\beta,(4,1)} \;&=\; 2\frac{g^2}{V^2}\sum_{\vk_1\to \vk_6}\delta_{\vk_1+\vk_6,\vk_2+\vk_5}\delta_{\vk_3+\vk_5,\vk_4+\vk_6}\sum_{\eta_5\eta_6}\{I_{1,\r}^{(4,1)}+I_{2,\r}^{(4,1)}\}
	\end{aligned}
\end{equation}
where
\begin{equation}
	\begin{aligned}
		I_{1, \r}^{(4,1)} &\equiv -\sum_{j_1j_2j_3j_4}\bar{\psi}_{0,K_1}\psi_{0,K_2}\bar{\psi}_{0,K_3}\psi_{0,K_4}\mathcal{T}_{\eta_5\eta_6}(\vk_5,\vk_6)[1-n_F(\eta_5M_5)]n_F(\eta_6M_6) \\
		&\mathrel{\phantom{=}}\; \times\; (n\beta)A_1(-\omega_{j_1}+\omega_{j_2},-\omega_{j_3}+\omega_{j_4},\eta_5M_5-\eta_6M_6)\\
		I_{2, \r}^{(4,1)} &\equiv -\sum_{j_1j_2j_3j_4}\bar{\psi}_{0,K_1}\psi_{0,K_2}\bar{\psi}_{0,K_3}\psi_{0,K_4}\mathcal{T}_{\eta_5\eta_6}(\vk_5,\vk_6)n_F(\eta_5M_5)[1-n_F(\eta_6M_6)] \\
		&\mathrel{\phantom{=}}\; \times\; (n\beta)A_2(-\omega_{j_1}+\omega_{j_2},-\omega_{j_3}+\omega_{j_4},-\eta_5M_5+\eta_6M_6)
	\end{aligned}
\end{equation}
The expression for $S_{\mu,\r}^{\beta,(4,2)}$ read,
\begin{equation}\label{eq-chap05-effective-action-replica-4-2}
	\begin{aligned}
		S_{\mu,\r}^{\beta,(4,2)} \;&=\;4g^2\frac{1}{V^2}\sum_{\vk_1\to\vk_6}\delta_{\vk_1+\vk_6,\vk_2+\vk_5}\delta_{\vk_3+\vk_5,\vk_4+\vk_6}\sum_{\eta_5\eta_6}\{I_{1,\r}^{(4,2)}+I_{2,\r}^{(4,2)}\}
	\end{aligned}
\end{equation}
where
\begin{equation}
	\begin{aligned}
		I_{1,\r}^{(4,2)} \;&=\; \sum_{j_1\to j_4}\bar{\psi}_{0,K_1}\psi_{0,K_2}\bar{\psi}_{0,K_3}\Lambda_{\vk_6}^{\eta_6}\gamma^0\Lambda_{\vk_5}^{\eta_5}\gamma^0\psi_{0,K_4}n_F(\eta_6M_6)[1-n_F(\eta_5M_5)] \\
        &\mathrel{\phantom{=}} \times\; (n\beta)A_1(-\omega_{j_1}+\omega_{j_2},-\omega_{j_3}+\omega_{j_4},\eta_5M_5-\eta_6M_6)\\
		I_{2,\r}^{(4,2)} \;&=\; \sum_{j_1\to j_4}\bar{\psi}_{0,K_1}\psi_{0,K_2}\bar{\psi}_{0,K_3}\Lambda_{\vk_6}^{\eta_6}\gamma^0\Lambda_{\vk_5}^{\eta_5}\gamma^0\psi_{0,K_4}n_F(\eta_5M_5)[1-n_F(\eta_6M_6)]\\
        &\mathrel{\phantom{=}} \times\; (n\beta)A_2(-\omega_{j_1}+\omega_{j_2},-\omega_{j_3}+\omega_{j_4},\eta_6M_6-\eta_5M_5)
	\end{aligned}
\end{equation}
The expression for $S_{\mu,\r}^{\beta,(4,3)}$ read,
\begin{equation}\label{eq-chap05-effective-action-replica-4-3}
	\begin{aligned}
		S_{\mu,\r}^{\beta,(4,3)} \;&=\; 2g^2\frac{1}{V^2}\sum_{\vk_1\to\vk_6}\delta_{\vk_1+\vk_6,\vk_2+\vk_5}\delta_{\vk_3+\vk_5,\vk_4+\vk_6}\sum_{\eta_5\eta_6}\{I_{1,\r}^{(4,3)}+I_{2,\r}^{(4,3)}\}
	\end{aligned}
\end{equation}
where
\begin{equation}
	\begin{aligned}
		I_{1,\r}^{(4,3)} \;&\equiv\; \sum_{j_1\to j_4}\bar{\psi}_{0,K_1}\Lambda_{\vk_5}^{\eta_5}\gamma^0\psi_{0,K_4}\bar{\psi}_{0,K_3}\Lambda_{\vk_6}^{\eta_6}\gamma^0\psi_{0,K_2}n_F(\eta_6M_6)[1-n_F(\eta_5M_5)]\\
        &\mathrel{\phantom{=}} \times\; (n\beta)A_1(-\omega_{j_1}+\omega_{j_2},-\omega_{j_3}+\omega_{j_4},\eta_5M_5-\eta_6M_6)\\
		I_{2,\r}^{(4,3)} \;&\equiv\; \sum_{j_1\to j_4}\bar{\psi}_{0,K_1}\Lambda_{\vk_5}^{\eta_5}\gamma^0\psi_{0,K_4}\bar{\psi}_{0,K_3}\Lambda_{\vk_6}^{\eta_6}\gamma^0\psi_{0,K_2}n_F(\eta_5M_5)[1-n_F(\eta_6M_6)]\\
        &\mathrel{\phantom{=}} \times\; (n\beta)A_2(-\omega_{j_1}+\omega_{j_2},-\omega_{j_3}+\omega_{j_4},\eta_6M_6-\eta_5M_5)
	\end{aligned}
\end{equation}
The expression for $S_{\mu,\r}^{\beta,(6)}$ read,
\begin{equation}\label{eq-chap05-effective-action-replica-6}
	\begin{aligned}
		S_{\mu,\r}^{\beta,(6)} \;&=\; -4g^2\frac{1}{V^2}\sum_{\vk_1\to \vk_7}\delta_{\vk_1+\vk_2,\vk_3+\vk_7}\cdot\delta_{\vk_7+\vk_4,\vk_5+\vk_6}\sum_{\eta_7}\{I_{1,\r}^{(6)}+I_{2,\r}^{(6)}\}\\
	\end{aligned}
\end{equation}
where
\begin{equation}
	\begin{aligned}
		I_{1,\r}^{(6)} &\equiv \sum_{j_1\to j_6}\bar{\psi}_{0,K_1}\psi_{0,K_3}\bar{\psi}_{0,K_4}\psi_{0,K_6}\bar{\psi}_{0,K_2}\Lambda_{\vk_7}^{\eta_7}\gamma^0\psi_{0,K_5}[1-n_F(\eta_7M_7)]\\
		&\mathrel{\phantom{=}} \times\; (n\beta)A_1(-\omega_{j_1}-\omega_{j_2}+\omega_{j_3},-\omega_{j_4}+\omega_{j_5}+\omega_{j_6}, \eta_7M_7) \\
		I_{2,\r}^{(6)} &\equiv -\sum_{j_1\to j_6}\bar{\psi}_{0,K_1}\psi_{0,K_3}\bar{\psi}_{0,K_4}\psi_{0,K_6}\bar{\psi}_{0,K_2}\Lambda_{\vk_7}^{\eta_7}\gamma^0\psi_{0,K_5}n_F(\eta_7M_7)\\
		&\mathrel{\phantom{=}} \times\; (n\beta)A_2(-\omega_{j_1}-\omega_{j_2}+\omega_{j_3},-\omega_{j_4}+\omega_{j_5}+\omega_{j_6}, -\eta_7M_7) \\
	\end{aligned}
\end{equation}

Recall that 
\begin{equation}
	\begin{aligned}
		&\mathrel{\phantom{=}} H_n(\mu) \\
		&= \frac{1}{n-1}\lim_{\beta\to\infty}\left(\braket{S_{\mu,r}^{n\beta,(2,1)}}+\braket{S_{\mu,r}^{n\beta,(2,2)}}+\braket{S_{\mu,r}^{n\beta,(4,1)}}+\braket{S_{\mu,r}^{n\beta,(4,2)}}+\braket{S_{\mu,r}^{n\beta,(4,3)}}+\braket{S_{\mu,r}^{n\beta,(6)}}\right. \\ 
		&\mathrel{\phantom{=}}\; \left.-\braket{S_{\mu}^{n\beta,(2,1)}}-\braket{S_{\mu}^{n\beta,(2,2)}}-\braket{S_{\mu}^{n\beta,(4,1)}}-\braket{S_{\mu}^{n\beta,(4,2)}}-\braket{S_{\mu}^{n\beta,(4,3)}}-\braket{S_{\mu}^{n\beta,(6)}}\right) 
	\end{aligned}
\end{equation}
now we prove the calim that the local part in "glued" effective action coincides with the oringinal effective action at temperture $n\beta$,
compare these two term, what we need to prove is actually the following identities,
\begin{equation}
	\begin{aligned}
		&\mathrel{\phantom{=}}\; \sum_{j_7}\delta_{j_1+j_2,j_3+j_7}\delta_{j_4+j_7,j_5+j_6}\S_0(K_7) \\
		&=\; \sum_{\eta_7}\left\{[1-n_F(\eta_7M_7)]B_1(-\omega_{j_1}-\omega_{j_2}+\omega_{j_3},-\omega_{j_4}+\omega_{j_5}+\omega_{j_6},\eta_7M_7)\right. \\
		&\mathrel{\phantom{=}}\;  +\; \left. n_F(\eta_7M_7)B_2(-\omega_{j_1}-\omega_{j_2}+\omega_{j_3},-\omega_{j_4}+\omega_{j_5}+\omega_{j_6},-\eta_7M_7)\right\}\Lambda_{\vk_7}^{\eta_7}\gamma^0\\
		&\mathrel{\phantom{=}}\; -\sum_{j_5,j_6}\delta_{j_1+j_6,j_2+j_5}\delta_{j_3+j_5,j_4+j_6}\S_0^{\eta_6}(K_6)\S_0^{\eta_5}(K_5)\\
		&=\; (n\beta)\left\{n_F(\eta_6M_6)[1-n_F(\eta_5M_5)]B_1(\omega_{j_2}-\omega_{j_1},\omega_{j_4}-\omega_{j_3},\eta_5M_5-\eta_6M_6)\right.\\
		&\mathrel{\phantom{=}}\; +\left.n_F(\eta_5M_5)[1-n_F(\eta_6M_6)]B_2(\omega_{j_2}-\omega_{j_1},\omega_{j_4}-\omega_{j_3},\eta_6M_6-\eta_5M_5)\right\}\\
		&\mathrel{\phantom{=}}\; \frac{1}{(n\beta)^2}\sum_{j_3j_4j_5}\delta_{j_1+j_5,j_3+j_4}\delta_{j_1,j_2}\S_0^{\eta_3}(K_3)\S_0^{\eta_4}(K_4)\S_0^{\eta_5}(K_5) \\
		&=\; -[1-n_F(\eta_3M_3)][1-n_F(\eta_4M_4)]n_F(\eta_5M_5)\times B_1(-\omega_{j_1},\omega_{j_2},\eta_3M_3+\eta_4M_4-\eta_5M_5) \\
		&\mathrel{\phantom{=}} \;+\; n_F(\eta_3M_3)n_F(\eta_4M_4)[1-n_F(\eta_5M_5)]\times B_2(-\omega_{j_1},\omega_{j_2},-\eta_3M_3-\eta_4M_4+\eta_5M_5)\\
	\end{aligned}
\end{equation}
These identities can be proved easily using the mixed representation of propagators.

Now we can write Rényi entropies as 
\begin{equation}
	H_n(\mu) \equiv H_n^{2,1}(\mu) + H_n^{2,2}(\mu) + H_n^{4,1}(\mu) + H_n^{4,2}(\mu) + H_n^{4,3}(\mu) + H_n^{6}(\mu)
\end{equation}
where
\begin{equation} 
	\begin{aligned}
		\mathrel{\phantom{=}} H_n^{2,1}(\mu) 
		&= \frac{1}{n-1}\lim_{\beta\to\infty}4g^2\frac{1}{V^2}\sum_{\vk_1\to \vk_5}\delta_{\vk_1+\vk_5,\vk_3+\vk_4}\delta_{\vk_1,\vk_2}\sum_{\eta_3\eta_4\eta_5}\sum_{j_1j_2}\braket{\bar{\psi}_{0,K_1}\Lambda_{\vk_4}^{\eta_4}\gamma^0\psi_{0,K_2}}\mathcal{T}_{\eta_3\eta_5}(\vk_3,\vk_5) \\
		&\mathrel{\phantom{=}} \times\; (n\beta)\left\{ n_F(\eta_3M_3)n_F(\eta_4M_4)[1-n_F(\eta_5M_5)]C_2(-\omega_{j_1},\omega_{j_2},-\eta_3M_3-\eta_4M_4+\eta_5M_5)  \right.\\
		&\mathrel{\phantom{=}}  -\; \left.[1-n_F(\eta_3M_3)][1-n_F(\eta_4M_4)]n_F(\eta_5M_5)C_1(-\omega_{j_1},\omega_{j_2},\eta_3M_3+\eta_4M_4-\eta_5M_5)   \right\}\\
	\end{aligned}
\end{equation}
\begin{equation} 
	\begin{aligned}
		\mathrel{\phantom{=}} H_n^{2,2}(\mu) 
		&= -\frac{1}{n-1}\lim_{\beta\to\infty}4g^2\frac{1}{V^2}\sum_{\vk_1\to \vk_5}\delta_{\vk_1+\vk_5,\vk_3+\vk_4}\delta_{\vk_1,\vk_2}\sum_{\eta_3\eta_4\eta_5}\sum_{j_1j_2}\braket{\bar{\psi}_{0,K_1}\Lambda_{\vk_4}^{\eta_4}\gamma^0\Lambda_{\vk_5}^{\eta_5}\gamma^0\Lambda_{\vk_3}^{\eta_3}\gamma^0\psi_{0,K_2}}\\
		&\mathrel{\phantom{=}} \times\; (n\beta)\left\{n_F(\eta_3M_3)n_F(\eta_4M_4)[1-n_F(\eta_5M_5)]C_2(-\omega_{j_1},\omega_{j_2},-\eta_3M_3-\eta_4M_4+\eta_5M_5)   \right.\\
		&\mathrel{\phantom{=}} -\; \left.  [1-n_F(\eta_3M_3)][1-n_F(\eta_4M_4)]n_F(\eta_5M_5)C_1(-\omega_{j_1},\omega_{j_2},\eta_3M_3+\eta_4M_4-\eta_5M_5) \right\}\\
	\end{aligned}
\end{equation}
\begin{equation}
	\begin{aligned}
		\mathrel{\phantom{=}} H_n^{4,1}(\mu) 
		&= \frac{1}{n-1}\lim_{\beta\to\infty}2g^2\frac{1}{V^2}\sum_{\vk_1\to \vk_6}\delta_{\vk_1+\vk_6,\vk_2+\vk_5}\delta_{\vk_3+\vk_5,\vk_4+\vk_6}\sum_{\eta_5\eta_6}\sum_{j_1\to j_4}\braket{\bar{\psi}_{0,K_1}\psi_{0,K_2}\bar{\psi}_{0,K_3}\psi_{0,K_4}}\\
		&\mathrel{\phantom{=}} \times\; \left\{ [1-n_F(\eta_5M_5)]n_F(\eta_6M_6)C_1(-\omega_{j_1}+\omega_{j_2},-\omega_{j_3}+\omega_{j_4},\eta_5M_5-\eta_6M_6) \right.\\
        &\mathrel{\phantom{=}} +\; \left.n_F(\eta_5M_5)[1-n_F(\eta_6M_6)]C_2(-\omega_{j_1}+\omega_{j_2},-\omega_{j_3}+\omega_{j_4},-\eta_5M_5+\eta_6M_6)\right\}\\
		&\mathrel{\phantom{=}} \times\; (-n\beta)\mathcal{T}_{\eta_5\eta_6}(\vk_5,\vk_6) \\
	\end{aligned}
\end{equation}
\begin{equation} 
	\begin{aligned}
		\mathrel{\phantom{=}} H_n^{4,2}(\mu) 
		&=\; \frac{1}{n-1}\lim_{\beta\to\infty}4g^2\frac{1}{V^2}\sum_{\vk_1\to\vk_6}\delta_{\vk_1+\vk_6,\vk_2+\vk_5}\delta_{\vk_3+\vk_5,\vk_4+\vk_6}\\
        &\mathrel{\phantom{=}} \times\; \sum_{\eta_5\eta_6}\sum_{j_1\to j_4}\braket{\bar{\psi}_{0,K_1}\psi_{0,K_2}\bar{\psi}_{0,K_3}\Lambda_{\vk_6}^{\eta_6}\gamma^0\Lambda_{\vk_5}^{\eta_5}\gamma^0\psi_{0,K_4}}(n\beta)\\
		&\mathrel{\phantom{=}} \times\;\left\{n_F(\eta_6M_6)[1-n_F(\eta_5M_5)]C_1(-\omega_{j_1}+\omega_{j_2},-\omega_{j_3}+\omega_{j_4},\eta_5M_5-\eta_6M_6)\right. \\
		&\mathrel{\phantom{=}} \left.+n_F(\eta_5M_5)[1-n_F(\eta_6M_6)]C_2(-\omega_{j_1}+\omega_{j_2},-\omega_{j_3}+\omega_{j_4},\eta_6M_6-\eta_5M_5)\right\} \\
	\end{aligned}
\end{equation}
\begin{equation}
	\begin{aligned}
		\mathrel{\phantom{=}} H_n^{4,3}(\mu) 
		&=\; \frac{1}{n-1}\lim_{\beta\to\infty}2g^2\frac{1}{V^2}\sum_{\vk_1\to\vk_6}\delta_{\vk_1+\vk_6,\vk_2+\vk_5}\delta_{\vk_3+\vk_5,\vk_4+\vk_6}\\
        &\mathrel{\phantom{=}} \times\; \sum_{\eta_5\eta_6}\sum_{j_1\to j_4}\braket{\bar{\psi}_{0,K_1}\Lambda_{\vk_5}^{\eta_5}\gamma^0\psi_{0,K_4}\bar{\psi}_{0,K_3}\Lambda_{\vk_6}^{\eta_6}\gamma^0\psi_{0,K_2}}(n\beta)\\
		&\mathrel{\phantom{=}} \times\; \left\{n_F(\eta_6M_6)[1-n_F(\eta_5M_5)]C_1(-\omega_{j_1}+\omega_{j_2},-\omega_{j_3}+\omega_{j_4},\eta_5M_5-\eta_6M_6)\right. \\
		&\mathrel{\phantom{=}} \left.+n_F(\eta_5M_5)[1-n_F(\eta_6M_6)]C_2(-\omega_{j_1}+\omega_{j_2},-\omega_{j_3}+\omega_{j_4},\eta_5M_5-\eta_6M_6)\right\} \\
	\end{aligned}
\end{equation}
\begin{equation}
	\begin{aligned}
		\mathrel{\phantom{=}} H_n^6(\mu) 
		&=\; \frac{1}{n-1}\lim_{\beta\to\infty}(-1)4g^2\frac{1}{V^2}\sum_{\vk_1\to \vk_7}\delta_{\vk_1+\vk_2,\vk_3+\vk_7}\cdot\delta_{\vk_7+\vk_4,\vk_5+\vk_6}\sum_{j_1\to j_6}\sum_{\eta_7} \\
		&\mathrel{\phantom{=}} \times\; \braket{\bar{\psi}_{0,K_1}\psi_{0,K_3}\bar{\psi}_{0,K_4}\psi_{0,K_6}\bar{\psi}_{0,K_2}\Lambda_{\vk_7}^{\eta_7}\gamma^0\psi_{0,K_5}}(n\beta)\\
		&\mathrel{\phantom{=}} \times\; \left\{ [1-n_F(\eta_7M_7)]\cdot C_1(-\omega_{j_1}-\omega_{j_2}+\omega_{j_3},-\omega_{j_4}+\omega_{j_5}+\omega_{j_6},\eta_7M_7) \right. \\
		&\mathrel{\phantom{=}} - \left. n_F(\eta_7M_7)C_2(-\omega_{j_1}-\omega_{j_2}+\omega_{j_3},-\omega_{j_4}+\omega_{j_5}+\omega_{j_6},-\eta_7M_7) \right\}
	\end{aligned}
\end{equation} 

For the remain calculation, we nee the following result,
\begin{equation}\label{eq-appendix-fermion-propagator-C-1}
	\begin{aligned}
		&\mathrel{\phantom{=}}\; \sum_{j_1,j_2}\delta_{j_1,j_2}\frac{1}{M_1+\i\omega_{j_1}}C_1(-\omega_{j_1},\omega_{j_2},M) \\
		&=\; n\cdot\left\{\frac{1-\e^{(M_1-M)\beta}}{(M-M_1)^2}\frac{1}{\e^{n\beta M_1}+1}-\frac{\beta}{M-M_1}\frac{1}{\e^{n\beta M}+1}\right\}
	\end{aligned}
\end{equation}

\begin{equation}\label{eq-appendix-fermion-propagator-C-2}
	\begin{aligned}
		&\mathrel{\phantom{=}}\; \sum_{j_1,j_2}\delta_{j_1,j_2}\frac{1}{M_1+\i\omega_{j_1}}C_2(-\omega_{j_1},\omega_{j_2},M)\\
		&=\; n\cdot \left\{\frac{\e^{-(M_1+M)\beta}-1}{(M+M_1)^2}\frac{\e^{n\beta M_1}}{\e^{n\beta M_1}+1}+\frac{\beta}{M+M_1}\frac{1}{\e^{n\beta M}+1}\right\}
	\end{aligned}
\end{equation}

For $H_n^{2,1}$, we have
\begin{equation*}
	\begin{aligned}
		&\mathrel{\phantom{=}}\; \braket{\bar{\psi}_{0,K_1}\Lambda_{\vk_4}^{\eta_4}\gamma^0\psi_{0,K_2}} = -\frac{1}{n\beta}\sum_{\eta_1}\mathcal{T}_{\eta_4\eta_1}(\vk_4,\vk_1)\S_0^{\eta_1}(K_1)\delta_{K_1,K_2}\\
	\end{aligned}
\end{equation*}
then we need to calculate 
\begin{equation}\label{eq-chap05-frequency-sum-2}
	\begin{aligned}
		\mathrel{\phantom{=}}\; \sum_{j_1j_2}\S_0^{\eta_1}(K_1)\delta_{j_1,j_2}
		&\times \left\{ -[1-n_F(\eta_3M_3)][1-n_F(\eta_4M_4)]n_F(\eta_5M_5)\cdot C_1(-\omega_{j_1},\omega_{j_2},\eta_3M_3+\eta_4M_4-\eta_5M_5) \right.\\
		&+\; \left.  n_F(\eta_3M_3)n_F(\eta_4M_4)[1-n_F(\eta_5M_5)]\cdot C_2(-\omega_{j_1},\omega_{j_2},-\eta_3M_3-\eta_4M_4+\eta_5M_5)  \right\}\\
	\end{aligned}
\end{equation}
Note that the factor $n_F(\eta M)$ give the constraint that $\eta = -1$ and $1-n_F(\eta M)$ gives $\eta = 1$ in order to survive the zero temperture limit,
substitute Eq.~\ref{eq-appendix-fermion-propagator-C-1} and Eq.~\ref{eq-appendix-fermion-propagator-C-2} we finally arrive at 
\begin{equation}
	\begin{aligned}
		\mathrel{\phantom{=}} H_n^{2,1}(\mu) &= \frac{n}{n-1}4g^2\frac{1}{V^2}\sum_{\vk_1\vk_3\vk_4 \vk_5}\delta_{\vk_1+\vk_5,\vk_3+\vk_4}\\
		&\mathrel{\phantom{=}} \times\; \frac{\mathcal{T}_{+-}(\vk_4,\vk_1)\mathcal{T}_{+-}(\vk_3,\vk_5)+\mathcal{T}_{-+}(\vk_4,\vk_1)\mathcal{T}_{-+}(\vk_3,\vk_5)}{(M_1+M_3+M_4+M_5)^2}
	\end{aligned}
\end{equation}

For For $H_n^{2,2}$, we have
\begin{equation}
	\begin{aligned}
		&\mathrel{\phantom{=}} \braket{\bar{\psi}_{0,K_1}\Lambda_{\vk_4}^{\eta_4}\gamma^0\Lambda_{\vk_5}^{\eta_5}\gamma^0\Lambda_{\vk_3}^{\eta_3}\gamma^0\psi_{0,K_2}}\\
		&=\; -\frac{1}{n\beta}\sum_{\eta_1}\text{tr}(\Lambda_{\vk_4}^{\eta_4}\gamma^0\Lambda_{\vk_5}^{\eta_5}\gamma^0\Lambda_{\vk_3}^{\eta_3}\gamma^0\Lambda_{\vk_1}^{\eta_1}\gamma^0)\S_0^{\eta_1}(K_1)\delta_{K_1,K_2}\\ 
	\end{aligned}
\end{equation}
For simplicity, define $\mathcal{T}_{\eta_1\eta_2\eta_3\eta_4}(\vk_1,\vk_2,\vk_3,\vk_4)\equiv\text{tr}(\Lambda_{\vk_1}^{\eta_1}\gamma^0\Lambda_{\vk_2}^{\eta_2}\gamma^0\Lambda_{\vk_3}^{\eta_3}\gamma^0\Lambda_{\vk_4}^{\eta_4}\gamma^0)$, according the result in 
Eq.~\ref{eq-chap05-frequency-sum-2} we get
\begin{equation}
	\begin{aligned}
		\mathrel{\phantom{=}} H_n^{2,2}(\mu) \;&=\; -\frac{n}{n-1}4g^2\frac{1}{V^2}\sum_{\vk_1\vk_3\vk_4 \vk_5}\delta_{\vk_1+\vk_5,\vk_3+\vk_4}\\
		&\mathrel{\phantom{=}} \times\; \frac{\mathcal{T}_{-+-+}(\vk_4,\vk_5,\vk_3,\vk_1)+\mathcal{T}_{+-+-}(\vk_4,\vk_5,\vk_3,\vk_1)}{(M_1+M_3+M_4+M_5)^2}\\ 
	\end{aligned}
\end{equation}

For $H_n^{4,1}$, 
\begin{equation}\label{eq-chap05-four-fermion-contraction}
	\begin{aligned}
		&\mathrel{\phantom{=}}\; \braket{\bar{\psi}_{0,K_1}\psi_{0,K_2}\bar{\psi}_{0,K_3}\psi_{0,K_4}} \\
		&=\; \braket{(\bar{\psi}_{0,K_1})_{\alpha}(\psi_{0,K_2})_{\alpha}(\bar{\psi}_{0,K_3})_{\beta}(\psi_{0,K_4})_\beta} \\
		&=\; \braket{(\psi_{0,K_2})_{\alpha}(\bar{\psi}_{0,K_1})_{\alpha}(\psi_{0,K_4})_\beta(\bar{\psi}_{0,K_3})_{\beta}} \\
		&=\; \frac{\S_0(K_1)_{\alpha\alpha}}{n\beta}\delta_{K_1,K_2}\frac{\S_0(K_3)_{\beta\beta}}{n\beta}\delta_{K_3,K_4}-\frac{\S_0(K_3)_{\alpha\beta}}{n\beta}\delta_{K_2,K_3}\frac{\S_0(K_1)_{\beta\alpha}}{n\beta}\delta_{K_4,K_1}
	\end{aligned}
\end{equation}
In this contraction we have two terms correspond to the two diagrams in Fig.~\ref{fig:contractiontype2}. We now show that one term contain $\delta_{K_1,K_2}\delta_{K_3,K_4}$ vanish 
even at finite temperture, this term is propotion to a factor as follows,
\begin{equation}
	\begin{aligned}
		&\mathrel{\phantom{=}} \left\{n_F(\eta_6M_6)[1-n_F(\eta_5M_5)]C_1(-\omega_{j_1}+\omega_{j_2},-\omega_{j_3}+\omega_{j_4},\eta_5M_5-\eta_6M_6)\right. \\
		&\mathrel{\phantom{=}} \left.+n_F(\eta_5M_5)[1-n_F(\eta_6M_6)]C_2(-\omega_{j_1}+\omega_{j_2},-\omega_{j_3}+\omega_{j_4},\eta_5M_5-\eta_6M_6)\right\} \\
	\end{aligned}
\end{equation}
given $\delta_{K_1K_2}\delta_{K_3,K_4}$,define $M\equiv \eta_5M_5-\eta_6M_6$, then we have 
\begin{equation}
	\begin{aligned}
		&\mathrel{\phantom{=}}\; n_F(\eta_6M_6)[1-n_F(\eta_5M_5)] C_1(0, 0, M) + n_F(\eta_5M_5)[1-n_F(\eta_6M_6)] C_2(0, 0, -M) \\
		&\propto\; n_F(\eta_6M_6)[1-n_F(\eta_5M_5)]\cdot \frac{e^{-M\beta}-1}{M^2} \;+\; n_F(\eta_5M_5[1-n_F(\eta_6M_6)]\cdot \frac{e^{M\beta}-1}{M^2} \\
		&\propto\; n_F(\eta_6M_6)[1-n_F(\eta_5M_5)]\Big\{ \frac{n_F(\eta_5M_5)}{1-n_F(\eta_5M_5)}\frac{1-n_F(\eta_6M_6)}{n_F(\eta_6M_6)} - 1\Big\} \\
		&\mathrel{\phantom{=}} +\; n_F(\eta_5M_5)[1-n_F(\eta_6M_6)]\cdot \Big\{ \frac{1-n_F(\eta_5M_5)}{n_F(\eta_5M_5)}\frac{n_F(\eta_6M_6)}{1-n_F(\eta_6M_6)} - 1\Big\} \\
		&=\; n_F(\eta_5M_5)[1-n_F(\eta_6M_6)]-n_F(\eta_6M_6)[1-n_F(\eta_5M_5)] \\
		&\mathrel{\phantom{=}} +\; n_F(\eta_6M_6)[1-n_F(\eta_5M_5)]-n_F(\eta_5M_5)[1-n_F(\eta_6M_6)] \\
		&=\; 0
	\end{aligned}
\end{equation}
The other term give a Matsubara summation as 
\begin{equation}\label{eq-chap05-two-propagator-C}
	\begin{aligned}
		&\sum_{j_1\to j_4}\delta_{j_1,j_4}\delta_{j_2,j_3}\S_0^{\eta_1}(K_1)\S_0^{\eta_3}(K_3)\\
		&\times\;\Big\{ n_F(\eta_6M_6)[1-n_F(\eta_5M_5)]\cdot C_1(-\omega_{j_1}+\omega_{j_2},-\omega_{j_3}+\omega_{j_4},\eta_5M_5-\eta_6M_6)\\
		&\mathrel{\phantom{=}}+\; n_F(\eta_5M_5)[1-n_F(\eta_6M_6)]\cdot C_2(-\omega_{j_1}+\omega_{j_2},-\omega_{j_3}+\omega_{j_4},-\eta_5M_5+\eta_6M_6) \Big\}
	\end{aligned}
\end{equation}
this summation involes a double usage of Eq.~\ref{eq-appendix-fermion-propagator-C-1} and Eq.~\ref{eq-appendix-fermion-propagator-C-2}, finally 
\begin{equation}
	\begin{aligned}
		H_n^{4,1}(\mu) \;&=\; \frac{n}{n-1}2g^2\frac{1}{V^2}\sum_{\vk_1\vk_3\vk_5\vk_6}\delta_{\vk_1+\vk_6,\vk_3+\vk_5}\\
		&\mathrel{\phantom{=}} \times\; \frac{\mathcal{T}_{-+}(\vk_1,\vk_3)\mathcal{T}_{+-}(\vk_5,\vk_6)+\mathcal{T}_{+-}(\vk_1,\vk_3)\mathcal{T}_{-+}(\vk_5,\vk_6)}{(M_1+M_3+M_5+M_6)^2} \\
	\end{aligned}
\end{equation}
similarly, we have
\begin{equation}
	\begin{aligned}
		H_n^{4,2}(\mu) \;&=\; -\frac{n}{n-1}4g^2\frac{1}{V^2}\sum_{\vk_1\vk_3\vk_5\vk_6}\delta_{\vk_1+\vk_6,\vk_3+\vk_5}\\
		&\mathrel{\phantom{=}}\;  \frac{\mathcal{T}_{-+-+}(\vk_3,\vk_6,\vk_5,\vk_1)+\mathcal{T}_{+-+-}(\vk_3,\vk_6,\vk_5,\vk_1)}{(M_1+M_3+M_5+M_6)^2} \\
		H_n^{4,3}(\mu) \;&=\; \frac{n}{n-1}2g^2\frac{1}{V^2}\sum_{\vk_1\vk_3\vk_5\vk_6}\delta_{\vk_1+\vk_6,\vk_3+\vk_5}\\
		&\mathrel{\phantom{=}} \times\; \frac{\mathcal{T}_{+-}(\vk_5,\vk_1)\mathcal{T}_{+-}(\vk_6,\vk_3)+\mathcal{T}_{-+}(\vk_5,\vk_1)\mathcal{T}_{-+}(\vk_6,\vk_3)}{(M_1+M_3+M_5+M_6)^2}\\
	\end{aligned}
\end{equation}
For $H_n^{6}$, we have a 6-point function,
there are six ways of contraction, four of them can't satify momentum conservation, only two of them are nonzero,
\begin{equation*}
	\begin{aligned}
		&\mathrel{\phantom{=}}\; \braket{\bar{\psi}_{0,K_1}\psi_{0,K_3}\bar{\psi}_{0,K_4}\psi_{0,K_6}\bar{\psi}_{0,K_2}\Lambda_{\vk_7}^{\eta_7}\gamma^0\psi_{0,K_5}}\\
		&\sim\; -(\Lambda_{\vk_7}^{\eta_7}\gamma^0)_{\rho\sigma}\left\{ -\; \frac{\S_0(K_4)_{\alpha\beta}}{n\beta}\delta_{K_3,K_4}\cdot\frac{\S_0(K_1)_{\beta\alpha}}{n\beta}\delta_{K_1,K_6}\cdot\frac{\S_0(K_2)_{\sigma\rho}}{n\beta}\delta_{K_5,K_2} \right.\\  
		&\mathrel{\phantom{=}}\; +\; \left. \frac{\S_0(K_4)_{\alpha\beta}}{n\beta}\delta_{K_3,K_4}\cdot\frac{\S_0(K_2)_{\beta\rho}}{n\beta}\delta_{K_2,K_6}\cdot\frac{\S_0(K_1)_{\sigma\alpha}}{n\beta}\delta_{K_5,K_1} \right\}\\ 
	\end{aligned}
\end{equation*}
we are left with a summation as 
\begin{equation}
	\begin{aligned}
		&\mathrel{\phantom{=}} \sum_{j_1\to j_6}\delta_{j_3,j_4}\delta_{j_1,j_6}\delta_{j_2,j_5}\S_0^{\eta_1}(K_1)\S_0^{\eta_2}(K_2)\S_0^{\eta_4}(K_4)  \\
		&\mathrel{\phantom{=}} \times\; \left\{ [1-n_F(\eta_7M_7)]\cdot C_1(-\omega_{j_1}-\omega_{j_2}+\omega_{j_3},-\omega_{j_4}+\omega_{j_5}+\omega_{j_6},\eta_7M_7) \right. \\
		&\mathrel{\phantom{=}} - \left. n_F(\eta_7M_7)C_2(-\omega_{j_1}-\omega_{j_2}+\omega_{j_3},-\omega_{j_4}+\omega_{j_5}+\omega_{j_6},-\eta_7M_7) \right\}
	\end{aligned}
\end{equation}
this summation can also be calculated by a treble usage of Eq.~\ref{eq-appendix-fermion-propagator-C-1} and Eq.~\ref{eq-appendix-fermion-propagator-C-2}, then we arrive at
\begin{equation}
	\begin{aligned}
		H_n^6(\mu) \;&=\; \frac{n}{n-1}4g^2\frac{1}{V^2}\sum_{\vk_1\vk_2\vk_3 \vk_7}\delta_{\vk_1+\vk_2,\vk_3+\vk_7}\frac{1}{(M_1+M_2+M_3+M_7)^2}\\
		&\mathrel{\phantom{=}} \times\; \Big\{\mathcal{T}_{+-}(\vk_3,\vk_1)\mathcal{T}_{+-}(\vk_7,\vk_2)-\mathcal{T}_{+-+-}(\vk_3,\vk_2,\vk_7,\vk_1) \\
		&\mathrel{\phantom{=}}\mathrel{\phantom{+}}\; +\; \mathcal{T}_{-+}(\vk_3,\vk_1)\mathcal{T}_{-+}(\vk_7,\vk_2)-\mathcal{T}_{-+-+}(\vk_3,\vk_2,\vk_7,\vk_1) \Big\}\\
	\end{aligned}
\end{equation}
 
\twocolumngrid 

\nocite{*}

\bibliography{apssamp}

\begin{thebibliography}{29}%
\makeatletter
\providecommand \@ifxundefined [1]{%
 \@ifx{#1\undefined}
}%
\providecommand \@ifnum [1]{%
 \ifnum #1\expandafter \@firstoftwo
 \else \expandafter \@secondoftwo
 \fi
}%
\providecommand \@ifx [1]{%
 \ifx #1\expandafter \@firstoftwo
 \else \expandafter \@secondoftwo
 \fi
}%
\providecommand \natexlab [1]{#1}%
\providecommand \enquote  [1]{``#1''}%
\providecommand \bibnamefont  [1]{#1}%
\providecommand \bibfnamefont [1]{#1}%
\providecommand \citenamefont [1]{#1}%
\providecommand \href@noop [0]{\@secondoftwo}%
\providecommand \href [0]{\begingroup \@sanitize@url \@href}%
\providecommand \@href[1]{\@@startlink{#1}\@@href}%
\providecommand \@@href[1]{\endgroup#1\@@endlink}%
\providecommand \@sanitize@url [0]{\catcode `\\12\catcode `\$12\catcode `\&12\catcode `\#12\catcode `\^12\catcode `\_12\catcode `\%12\relax}%
\providecommand \@@startlink[1]{}%
\providecommand \@@endlink[0]{}%
\providecommand \url  [0]{\begingroup\@sanitize@url \@url }%
\providecommand \@url [1]{\endgroup\@href {#1}{\urlprefix }}%
\providecommand \urlprefix  [0]{URL }%
\providecommand \Eprint [0]{\href }%
\providecommand \doibase [0]{https://doi.org/}%
\providecommand \selectlanguage [0]{\@gobble}%
\providecommand \bibinfo  [0]{\@secondoftwo}%
\providecommand \bibfield  [0]{\@secondoftwo}%
\providecommand \translation [1]{[#1]}%
\providecommand \BibitemOpen [0]{}%
\providecommand \bibitemStop [0]{}%
\providecommand \bibitemNoStop [0]{.\EOS\space}%
\providecommand \EOS [0]{\spacefactor3000\relax}%
\providecommand \BibitemShut  [1]{\csname bibitem#1\endcsname}%
\let\auto@bib@innerbib\@empty
\bibitem [{\citenamefont {Michael A.~Nielsen}(2010)}]{RN196}%
  \BibitemOpen
  \bibfield  {author} {\bibinfo {author} {\bibfnamefont {I.~L.~C.}\ \bibnamefont {Michael A.~Nielsen}},\ }\href {https://doi.org/https://doi.org/10.1017/CBO9780511976667} {\emph {\bibinfo {title} {Quantum Computation and Quantum Information}}}\ (\bibinfo {year} {2010})\BibitemShut {NoStop}%
\bibitem [{\citenamefont {Bombelli}\ \emph {et~al.}(1986)\citenamefont {Bombelli}, \citenamefont {Koul}, \citenamefont {Lee},\ and\ \citenamefont {Sorkin}}]{RN116}%
  \BibitemOpen
  \bibfield  {author} {\bibinfo {author} {\bibfnamefont {L.}~\bibnamefont {Bombelli}}, \bibinfo {author} {\bibfnamefont {R.~K.}\ \bibnamefont {Koul}}, \bibinfo {author} {\bibfnamefont {J.}~\bibnamefont {Lee}},\ and\ \bibinfo {author} {\bibfnamefont {R.~D.}\ \bibnamefont {Sorkin}},\ }\bibfield  {title} {\bibinfo {title} {Quantum source of entropy for black holes},\ }\href {https://doi.org/10.1103/physrevd.34.373} {\bibfield  {journal} {\bibinfo  {journal} {Phys Rev D Part Fields}\ }\textbf {\bibinfo {volume} {34}},\ \bibinfo {pages} {373} (\bibinfo {year} {1986})},\ \bibinfo {note} {bombelli Koul Lee Sorkin eng Phys Rev D Part Fields. 1986 Jul 15;34(2):373-383. doi: 10.1103/physrevd.34.373.}\BibitemShut {Stop}%
\bibitem [{\citenamefont {Callan}\ and\ \citenamefont {Wilczek}(1994)}]{RN119}%
  \BibitemOpen
  \bibfield  {author} {\bibinfo {author} {\bibfnamefont {C.}~\bibnamefont {Callan}}\ and\ \bibinfo {author} {\bibfnamefont {F.}~\bibnamefont {Wilczek}},\ }\bibfield  {title} {\bibinfo {title} {On geometric entropy},\ }\href {https://doi.org/10.1016/0370-2693(94)91007-3} {\bibfield  {journal} {\bibinfo  {journal} {Physics Letters B}\ }\textbf {\bibinfo {volume} {333}},\ \bibinfo {pages} {55} (\bibinfo {year} {1994})}\BibitemShut {NoStop}%
\bibitem [{\citenamefont {Srednicki}(1993)}]{RN115}%
  \BibitemOpen
  \bibfield  {author} {\bibinfo {author} {\bibfnamefont {M.}~\bibnamefont {Srednicki}},\ }\bibfield  {title} {\bibinfo {title} {Entropy and area},\ }\href {https://doi.org/10.1103/PhysRevLett.71.666} {\bibfield  {journal} {\bibinfo  {journal} {Phys Rev Lett}\ }\textbf {\bibinfo {volume} {71}},\ \bibinfo {pages} {666} (\bibinfo {year} {1993})},\ \bibinfo {note} {srednicki eng Phys Rev Lett. 1993 Aug 2;71(5):666-669. doi: 10.1103/PhysRevLett.71.666.}\BibitemShut {Stop}%
\bibitem [{\citenamefont {Ryu}\ and\ \citenamefont {Takayanagi}(2006)}]{RN49}%
  \BibitemOpen
  \bibfield  {author} {\bibinfo {author} {\bibfnamefont {S.}~\bibnamefont {Ryu}}\ and\ \bibinfo {author} {\bibfnamefont {T.}~\bibnamefont {Takayanagi}},\ }\bibfield  {title} {\bibinfo {title} {Holographic derivation of entanglement entropy from the anti-de sitter space/conformal field theory correspondence},\ }\href {https://doi.org/10.1103/PhysRevLett.96.181602} {\bibfield  {journal} {\bibinfo  {journal} {Phys Rev Lett}\ }\textbf {\bibinfo {volume} {96}},\ \bibinfo {pages} {181602} (\bibinfo {year} {2006})},\ \bibinfo {note} {ryu, Shinsei Takayanagi, Tadashi eng 2006/05/23 Phys Rev Lett. 2006 May 12;96(18):181602. doi: 10.1103/PhysRevLett.96.181602. Epub 2006 May 9.}\BibitemShut {Stop}%
\bibitem [{\citenamefont {Kitaev}\ and\ \citenamefont {Preskill}(2006)}]{RN99}%
  \BibitemOpen
  \bibfield  {author} {\bibinfo {author} {\bibfnamefont {A.}~\bibnamefont {Kitaev}}\ and\ \bibinfo {author} {\bibfnamefont {J.}~\bibnamefont {Preskill}},\ }\bibfield  {title} {\bibinfo {title} {Topological entanglement entropy},\ }\href {https://doi.org/10.1103/PhysRevLett.96.110404} {\bibfield  {journal} {\bibinfo  {journal} {Phys Rev Lett}\ }\textbf {\bibinfo {volume} {96}},\ \bibinfo {pages} {110404} (\bibinfo {year} {2006})},\ \bibinfo {note} {kitaev, Alexei Preskill, John eng 2006/04/12 Phys Rev Lett. 2006 Mar 24;96(11):110404. doi: 10.1103/PhysRevLett.96.110404. Epub 2006 Mar 24.}\BibitemShut {Stop}%
\bibitem [{\citenamefont {Levin}\ and\ \citenamefont {Wen}(2006)}]{RN100}%
  \BibitemOpen
  \bibfield  {author} {\bibinfo {author} {\bibfnamefont {M.}~\bibnamefont {Levin}}\ and\ \bibinfo {author} {\bibfnamefont {X.~G.}\ \bibnamefont {Wen}},\ }\bibfield  {title} {\bibinfo {title} {Detecting topological order in a ground state wave function},\ }\href {https://doi.org/10.1103/PhysRevLett.96.110405} {\bibfield  {journal} {\bibinfo  {journal} {Phys Rev Lett}\ }\textbf {\bibinfo {volume} {96}},\ \bibinfo {pages} {110405} (\bibinfo {year} {2006})},\ \bibinfo {note} {levin, Michael Wen, Xiao-Gang eng 2006/04/12 Phys Rev Lett. 2006 Mar 24;96(11):110405. doi: 10.1103/PhysRevLett.96.110405. Epub 2006 Mar 24.}\BibitemShut {Stop}%
\bibitem [{\citenamefont {Shi}(2004)}]{RN195}%
  \BibitemOpen
  \bibfield  {author} {\bibinfo {author} {\bibfnamefont {Y.}~\bibnamefont {Shi}},\ }\bibfield  {title} {\bibinfo {title} {Entanglement in relativistic quantum field theory},\ }\bibfield  {journal} {\bibinfo  {journal} {Physical Review D}\ }\textbf {\bibinfo {volume} {70}},\ \href {https://doi.org/10.1103/PhysRevD.70.105001} {10.1103/PhysRevD.70.105001} (\bibinfo {year} {2004})\BibitemShut {NoStop}%
\bibitem [{\citenamefont {Casini}\ and\ \citenamefont {Huerta}(2009)}]{RN112}%
  \BibitemOpen
  \bibfield  {author} {\bibinfo {author} {\bibfnamefont {H.}~\bibnamefont {Casini}}\ and\ \bibinfo {author} {\bibfnamefont {M.}~\bibnamefont {Huerta}},\ }\bibfield  {title} {\bibinfo {title} {Entanglement entropy in free quantum field theory},\ }\bibfield  {journal} {\bibinfo  {journal} {Journal of Physics A: Mathematical and Theoretical}\ }\textbf {\bibinfo {volume} {42}},\ \href {https://doi.org/10.1088/1751-8113/42/50/504007} {10.1088/1751-8113/42/50/504007} (\bibinfo {year} {2009})\BibitemShut {NoStop}%
\bibitem [{\citenamefont {Calabrese}\ and\ \citenamefont {Cardy}(2004)}]{RN117}%
  \BibitemOpen
  \bibfield  {author} {\bibinfo {author} {\bibfnamefont {P.}~\bibnamefont {Calabrese}}\ and\ \bibinfo {author} {\bibfnamefont {J.}~\bibnamefont {Cardy}},\ }\bibfield  {title} {\bibinfo {title} {Entanglement entropy and quantum field theory},\ }\bibfield  {journal} {\bibinfo  {journal} {Journal of Statistical Mechanics: Theory and Experiment}\ }\textbf {\bibinfo {volume} {2004}},\ \href {https://doi.org/10.1088/1742-5468/2004/06/p06002} {10.1088/1742-5468/2004/06/p06002} (\bibinfo {year} {2004})\BibitemShut {NoStop}%
\bibitem [{\citenamefont {Calabrese}\ and\ \citenamefont {Cardy}(2009)}]{RN80}%
  \BibitemOpen
  \bibfield  {author} {\bibinfo {author} {\bibfnamefont {P.}~\bibnamefont {Calabrese}}\ and\ \bibinfo {author} {\bibfnamefont {J.}~\bibnamefont {Cardy}},\ }\bibfield  {title} {\bibinfo {title} {Entanglement entropy and conformal field theory},\ }\bibfield  {journal} {\bibinfo  {journal} {Journal of Physics A: Mathematical and Theoretical}\ }\textbf {\bibinfo {volume} {42}},\ \href {https://doi.org/10.1088/1751-8113/42/50/504005} {10.1088/1751-8113/42/50/504005} (\bibinfo {year} {2009})\BibitemShut {NoStop}%
\bibitem [{\citenamefont {Horodecki}\ \emph {et~al.}(2009)\citenamefont {Horodecki}, \citenamefont {Horodecki}, \citenamefont {Horodecki},\ and\ \citenamefont {Horodecki}}]{RN47}%
  \BibitemOpen
  \bibfield  {author} {\bibinfo {author} {\bibfnamefont {R.}~\bibnamefont {Horodecki}}, \bibinfo {author} {\bibfnamefont {P.}~\bibnamefont {Horodecki}}, \bibinfo {author} {\bibfnamefont {M.}~\bibnamefont {Horodecki}},\ and\ \bibinfo {author} {\bibfnamefont {K.}~\bibnamefont {Horodecki}},\ }\bibfield  {title} {\bibinfo {title} {Quantum entanglement},\ }\href {https://doi.org/10.1103/RevModPhys.81.865} {\bibfield  {journal} {\bibinfo  {journal} {Reviews of Modern Physics}\ }\textbf {\bibinfo {volume} {81}},\ \bibinfo {pages} {865} (\bibinfo {year} {2009})}\BibitemShut {NoStop}%
\bibitem [{\citenamefont {Nishioka}(2018)}]{RN120}%
  \BibitemOpen
  \bibfield  {author} {\bibinfo {author} {\bibfnamefont {T.}~\bibnamefont {Nishioka}},\ }\bibfield  {title} {\bibinfo {title} {Entanglement entropy: Holography and renormalization group},\ }\bibfield  {journal} {\bibinfo  {journal} {Reviews of Modern Physics}\ }\textbf {\bibinfo {volume} {90}},\ \href {https://doi.org/10.1103/RevModPhys.90.035007} {10.1103/RevModPhys.90.035007} (\bibinfo {year} {2018})\BibitemShut {NoStop}%
\bibitem [{\citenamefont {Witten}(2018)}]{RN113}%
  \BibitemOpen
  \bibfield  {author} {\bibinfo {author} {\bibfnamefont {E.}~\bibnamefont {Witten}},\ }\bibfield  {title} {\bibinfo {title} {Aps medal for exceptional achievement in research: Invited article on entanglement properties of quantum field theory},\ }\bibfield  {journal} {\bibinfo  {journal} {Reviews of Modern Physics}\ }\textbf {\bibinfo {volume} {90}},\ \href {https://doi.org/10.1103/RevModPhys.90.045003} {10.1103/RevModPhys.90.045003} (\bibinfo {year} {2018})\BibitemShut {NoStop}%
\bibitem [{\citenamefont {Balasubramanian}\ \emph {et~al.}(2012)\citenamefont {Balasubramanian}, \citenamefont {McDermott},\ and\ \citenamefont {Van~Raamsdonk}}]{RN121}%
  \BibitemOpen
  \bibfield  {author} {\bibinfo {author} {\bibfnamefont {V.}~\bibnamefont {Balasubramanian}}, \bibinfo {author} {\bibfnamefont {M.~B.}\ \bibnamefont {McDermott}},\ and\ \bibinfo {author} {\bibfnamefont {M.}~\bibnamefont {Van~Raamsdonk}},\ }\bibfield  {title} {\bibinfo {title} {Momentum-space entanglement and renormalization in quantum field theory},\ }\bibfield  {journal} {\bibinfo  {journal} {Physical Review D}\ }\textbf {\bibinfo {volume} {86}},\ \href {https://doi.org/10.1103/PhysRevD.86.045014} {10.1103/PhysRevD.86.045014} (\bibinfo {year} {2012})\BibitemShut {NoStop}%
\bibitem [{\citenamefont {Martins~Costa}\ \emph {et~al.}(2022)\citenamefont {Martins~Costa}, \citenamefont {van~den Brink}, \citenamefont {Nogueira},\ and\ \citenamefont {Krein}}]{RN65}%
  \BibitemOpen
  \bibfield  {author} {\bibinfo {author} {\bibfnamefont {M.~H.}\ \bibnamefont {Martins~Costa}}, \bibinfo {author} {\bibfnamefont {J.}~\bibnamefont {van~den Brink}}, \bibinfo {author} {\bibfnamefont {F.~S.}\ \bibnamefont {Nogueira}},\ and\ \bibinfo {author} {\bibfnamefont {G.~I.}\ \bibnamefont {Krein}},\ }\bibfield  {title} {\bibinfo {title} {Momentum space entanglement from the wilsonian effective action},\ }\bibfield  {journal} {\bibinfo  {journal} {Physical Review D}\ }\textbf {\bibinfo {volume} {106}},\ \href {https://doi.org/10.1103/PhysRevD.106.065024} {10.1103/PhysRevD.106.065024} (\bibinfo {year} {2022})\BibitemShut {NoStop}%
\bibitem [{\citenamefont {Martins~Costa}\ \emph {et~al.}(2023)\citenamefont {Martins~Costa}, \citenamefont {van~den Brink}, \citenamefont {Nogueira},\ and\ \citenamefont {Krein}}]{RN108}%
  \BibitemOpen
  \bibfield  {author} {\bibinfo {author} {\bibfnamefont {M.~H.}\ \bibnamefont {Martins~Costa}}, \bibinfo {author} {\bibfnamefont {J.}~\bibnamefont {van~den Brink}}, \bibinfo {author} {\bibfnamefont {F.~S.}\ \bibnamefont {Nogueira}},\ and\ \bibinfo {author} {\bibfnamefont {G.~I.}\ \bibnamefont {Krein}},\ }\bibfield  {title} {\bibinfo {title} {Wilsonian renormalization as a quantum channel and the separability of fixed points},\ }\bibfield  {journal} {\bibinfo  {journal} {Physical Review D}\ }\textbf {\bibinfo {volume} {107}},\ \href {https://doi.org/10.1103/PhysRevD.107.125014} {10.1103/PhysRevD.107.125014} (\bibinfo {year} {2023})\BibitemShut {NoStop}%
\bibitem [{\citenamefont {Hsu}\ \emph {et~al.}(2013{\natexlab{a}})\citenamefont {Hsu}, \citenamefont {McDermott},\ and\ \citenamefont {Van~Raamsdonk}}]{RN72}%
  \BibitemOpen
  \bibfield  {author} {\bibinfo {author} {\bibfnamefont {T.-C.~L.}\ \bibnamefont {Hsu}}, \bibinfo {author} {\bibfnamefont {M.~B.}\ \bibnamefont {McDermott}},\ and\ \bibinfo {author} {\bibfnamefont {M.}~\bibnamefont {Van~Raamsdonk}},\ }\bibfield  {title} {\bibinfo {title} {Momentum-space entanglement for interacting fermions at finite density},\ }\bibfield  {journal} {\bibinfo  {journal} {Journal of High Energy Physics}\ }\textbf {\bibinfo {volume} {2013}},\ \href {https://doi.org/10.1007/jhep11(2013)121} {10.1007/jhep11(2013)121} (\bibinfo {year} {2013}{\natexlab{a}})\BibitemShut {NoStop}%
\bibitem [{\citenamefont {Brahma}\ \emph {et~al.}(2023)\citenamefont {Brahma}, \citenamefont {Calderón-Figueroa}, \citenamefont {Hassan},\ and\ \citenamefont {Mi}}]{RN87}%
  \BibitemOpen
  \bibfield  {author} {\bibinfo {author} {\bibfnamefont {S.}~\bibnamefont {Brahma}}, \bibinfo {author} {\bibfnamefont {J.}~\bibnamefont {Calderón-Figueroa}}, \bibinfo {author} {\bibfnamefont {M.}~\bibnamefont {Hassan}},\ and\ \bibinfo {author} {\bibfnamefont {X.}~\bibnamefont {Mi}},\ }\bibfield  {title} {\bibinfo {title} {Momentum-space entanglement entropy in de sitter spacetime},\ }\bibfield  {journal} {\bibinfo  {journal} {Physical Review D}\ }\textbf {\bibinfo {volume} {108}},\ \href {https://doi.org/10.1103/PhysRevD.108.043522} {10.1103/PhysRevD.108.043522} (\bibinfo {year} {2023})\BibitemShut {NoStop}%
\bibitem [{\citenamefont {Burgess}\ \emph {et~al.}(2025)\citenamefont {Burgess}, \citenamefont {Colas}, \citenamefont {Holman},\ and\ \citenamefont {Kaplanek}}]{RN167}%
  \BibitemOpen
  \bibfield  {author} {\bibinfo {author} {\bibfnamefont {C.~P.}\ \bibnamefont {Burgess}}, \bibinfo {author} {\bibfnamefont {T.}~\bibnamefont {Colas}}, \bibinfo {author} {\bibfnamefont {R.}~\bibnamefont {Holman}},\ and\ \bibinfo {author} {\bibfnamefont {G.}~\bibnamefont {Kaplanek}},\ }\bibfield  {title} {\bibinfo {title} {Does decoherence violate decoupling?},\ }\bibfield  {journal} {\bibinfo  {journal} {Journal of High Energy Physics}\ }\textbf {\bibinfo {volume} {2025}},\ \href {https://doi.org/10.1007/jhep02(2025)204} {10.1007/jhep02(2025)204} (\bibinfo {year} {2025})\BibitemShut {NoStop}%
\bibitem [{\citenamefont {Flynn}\ \emph {et~al.}(2023)\citenamefont {Flynn}, \citenamefont {Tang}, \citenamefont {Chandran},\ and\ \citenamefont {Laumann}}]{RN86}%
  \BibitemOpen
  \bibfield  {author} {\bibinfo {author} {\bibfnamefont {M.~O.}\ \bibnamefont {Flynn}}, \bibinfo {author} {\bibfnamefont {L.-H.}\ \bibnamefont {Tang}}, \bibinfo {author} {\bibfnamefont {A.}~\bibnamefont {Chandran}},\ and\ \bibinfo {author} {\bibfnamefont {C.~R.}\ \bibnamefont {Laumann}},\ }\bibfield  {title} {\bibinfo {title} {Momentum space entanglement of interacting fermions},\ }\bibfield  {journal} {\bibinfo  {journal} {Physical Review B}\ }\textbf {\bibinfo {volume} {107}},\ \href {https://doi.org/10.1103/PhysRevB.107.L081109} {10.1103/PhysRevB.107.L081109} (\bibinfo {year} {2023})\BibitemShut {NoStop}%
\bibitem [{\citenamefont {Hsu}\ \emph {et~al.}(2013{\natexlab{b}})\citenamefont {Hsu}, \citenamefont {McDermott},\ and\ \citenamefont {Van~Raamsdonk}}]{RRN72}%
  \BibitemOpen
  \bibfield  {author} {\bibinfo {author} {\bibfnamefont {T.-C.~L.}\ \bibnamefont {Hsu}}, \bibinfo {author} {\bibfnamefont {M.~B.}\ \bibnamefont {McDermott}},\ and\ \bibinfo {author} {\bibfnamefont {M.}~\bibnamefont {Van~Raamsdonk}},\ }\bibfield  {title} {\bibinfo {title} {Momentum-space entanglement for interacting fermions at finite density},\ }\bibfield  {journal} {\bibinfo  {journal} {Journal of High Energy Physics}\ }\textbf {\bibinfo {volume} {2013}},\ \href {https://doi.org/10.1007/jhep11(2013)121} {10.1007/jhep11(2013)121} (\bibinfo {year} {2013}{\natexlab{b}})\BibitemShut {NoStop}%
\bibitem [{\citenamefont {Lundgren}\ \emph {et~al.}(2019)\citenamefont {Lundgren}, \citenamefont {Liu}, \citenamefont {Laurell},\ and\ \citenamefont {Fiete}}]{RN197}%
  \BibitemOpen
  \bibfield  {author} {\bibinfo {author} {\bibfnamefont {R.}~\bibnamefont {Lundgren}}, \bibinfo {author} {\bibfnamefont {F.}~\bibnamefont {Liu}}, \bibinfo {author} {\bibfnamefont {P.}~\bibnamefont {Laurell}},\ and\ \bibinfo {author} {\bibfnamefont {G.~A.}\ \bibnamefont {Fiete}},\ }\bibfield  {title} {\bibinfo {title} {Momentum-space entanglement after a quench in one-dimensional disordered fermionic systems},\ }\bibfield  {journal} {\bibinfo  {journal} {Physical Review B}\ }\textbf {\bibinfo {volume} {100}},\ \href {https://doi.org/10.1103/PhysRevB.100.241108} {10.1103/PhysRevB.100.241108} (\bibinfo {year} {2019})\BibitemShut {NoStop}%
\bibitem [{\citenamefont {Mondragon-Shem}\ \emph {et~al.}(2013)\citenamefont {Mondragon-Shem}, \citenamefont {Khan},\ and\ \citenamefont {Hughes}}]{RN198}%
  \BibitemOpen
  \bibfield  {author} {\bibinfo {author} {\bibfnamefont {I.}~\bibnamefont {Mondragon-Shem}}, \bibinfo {author} {\bibfnamefont {M.}~\bibnamefont {Khan}},\ and\ \bibinfo {author} {\bibfnamefont {T.~L.}\ \bibnamefont {Hughes}},\ }\bibfield  {title} {\bibinfo {title} {Characterizing disordered fermion systems using the momentum-space entanglement spectrum},\ }\href {https://doi.org/10.1103/PhysRevLett.110.046806} {\bibfield  {journal} {\bibinfo  {journal} {Phys Rev Lett}\ }\textbf {\bibinfo {volume} {110}},\ \bibinfo {pages} {046806} (\bibinfo {year} {2013})},\ \bibinfo {note} {mondragon-Shem, Ian Khan, Mayukh Hughes, Taylor L eng 2013/01/25 Phys Rev Lett. 2013 Jan 25;110(4):046806. doi: 10.1103/PhysRevLett.110.046806. Epub 2013 Jan 25.}\BibitemShut {Stop}%
\bibitem [{\citenamefont {Torlai}\ \emph {et~al.}(2014)\citenamefont {Torlai}, \citenamefont {Tagliacozzo},\ and\ \citenamefont {De~Chiara}}]{RN199}%
  \BibitemOpen
  \bibfield  {author} {\bibinfo {author} {\bibfnamefont {G.}~\bibnamefont {Torlai}}, \bibinfo {author} {\bibfnamefont {L.}~\bibnamefont {Tagliacozzo}},\ and\ \bibinfo {author} {\bibfnamefont {G.}~\bibnamefont {De~Chiara}},\ }\bibfield  {title} {\bibinfo {title} {Dynamics of the entanglement spectrum in spin chains},\ }\bibfield  {journal} {\bibinfo  {journal} {Journal of Statistical Mechanics: Theory and Experiment}\ }\textbf {\bibinfo {volume} {2014}},\ \href {https://doi.org/10.1088/1742-5468/2014/06/p06001} {10.1088/1742-5468/2014/06/p06001} (\bibinfo {year} {2014})\BibitemShut {NoStop}%
\bibitem [{\citenamefont {Lundgren}\ \emph {et~al.}(2014)\citenamefont {Lundgren}, \citenamefont {Blair}, \citenamefont {Greiter}, \citenamefont {Lauchli}, \citenamefont {Fiete},\ and\ \citenamefont {Thomale}}]{RN200}%
  \BibitemOpen
  \bibfield  {author} {\bibinfo {author} {\bibfnamefont {R.}~\bibnamefont {Lundgren}}, \bibinfo {author} {\bibfnamefont {J.}~\bibnamefont {Blair}}, \bibinfo {author} {\bibfnamefont {M.}~\bibnamefont {Greiter}}, \bibinfo {author} {\bibfnamefont {A.}~\bibnamefont {Lauchli}}, \bibinfo {author} {\bibfnamefont {G.~A.}\ \bibnamefont {Fiete}},\ and\ \bibinfo {author} {\bibfnamefont {R.}~\bibnamefont {Thomale}},\ }\bibfield  {title} {\bibinfo {title} {Momentum-space entanglement spectrum of bosons and fermions with interactions},\ }\href {https://doi.org/10.1103/PhysRevLett.113.256404} {\bibfield  {journal} {\bibinfo  {journal} {Phys Rev Lett}\ }\textbf {\bibinfo {volume} {113}},\ \bibinfo {pages} {256404} (\bibinfo {year} {2014})},\ \bibinfo {note} {lundgren, Rex Blair, Jonathan Greiter, Martin Lauchli, Andreas Fiete, Gregory A Thomale, Ronny eng 2015/01/03 Phys Rev Lett. 2014 Dec 19;113(25):256404. doi: 10.1103/PhysRevLett.113.256404. Epub 2014 Dec 18.}\BibitemShut {Stop}%
\bibitem [{\citenamefont {Ibáñez-Berganza}\ \emph {et~al.}(2016)\citenamefont {Ibáñez-Berganza}, \citenamefont {Rodríguez-Laguna},\ and\ \citenamefont {Sierra}}]{RN201}%
  \BibitemOpen
  \bibfield  {author} {\bibinfo {author} {\bibfnamefont {M.}~\bibnamefont {Ibáñez-Berganza}}, \bibinfo {author} {\bibfnamefont {J.}~\bibnamefont {Rodríguez-Laguna}},\ and\ \bibinfo {author} {\bibfnamefont {G.}~\bibnamefont {Sierra}},\ }\bibfield  {title} {\bibinfo {title} {Fourier-space entanglement of spin chains},\ }\bibfield  {journal} {\bibinfo  {journal} {Journal of Statistical Mechanics: Theory and Experiment}\ }\textbf {\bibinfo {volume} {2016}},\ \href {https://doi.org/10.1088/1742-5468/2016/05/053112} {10.1088/1742-5468/2016/05/053112} (\bibinfo {year} {2016})\BibitemShut {NoStop}%
\bibitem [{\citenamefont {Wilson}(1971{\natexlab{a}})}]{RN66}%
  \BibitemOpen
  \bibfield  {author} {\bibinfo {author} {\bibfnamefont {K.~G.}\ \bibnamefont {Wilson}},\ }\bibfield  {title} {\bibinfo {title} {Renormalization group and critical phenomena. i. renormalization group and the kadanoff scaling picture},\ }\href {https://doi.org/10.1103/PhysRevB.4.3174} {\bibfield  {journal} {\bibinfo  {journal} {Physical Review B}\ }\textbf {\bibinfo {volume} {4}},\ \bibinfo {pages} {3174} (\bibinfo {year} {1971}{\natexlab{a}})}\BibitemShut {NoStop}%
\bibitem [{\citenamefont {Wilson}(1971{\natexlab{b}})}]{RN67}%
  \BibitemOpen
  \bibfield  {author} {\bibinfo {author} {\bibfnamefont {K.~G.}\ \bibnamefont {Wilson}},\ }\bibfield  {title} {\bibinfo {title} {Renormalization group and critical phenomena. ii. phase-space cell analysis of critical behavior},\ }\href {https://doi.org/10.1103/PhysRevB.4.3184} {\bibfield  {journal} {\bibinfo  {journal} {Physical Review B}\ }\textbf {\bibinfo {volume} {4}},\ \bibinfo {pages} {3184} (\bibinfo {year} {1971}{\natexlab{b}})}\BibitemShut {NoStop}%
\end{thebibliography}%

\end{document}